\RequirePackage{fix-cm}
\PassOptionsToPackage{table, dvipsnames, prologue}{xcolor}
\documentclass[smallcondensed]{svjour3}       
\usepackage{graphicx}
\usepackage{booktabs, nicematrix}
\usepackage{lscape}
\usepackage{fontawesome}
\usepackage{tabularx}
\usepackage{tabularray}
\usepackage{pdfpages}
\usepackage{soul}
\usepackage{url}
\usepackage{dirtytalk}
\usepackage{natbib}
\usepackage{stix}
\usepackage{tikz}
\usepackage{marginnote}
\usepackage{dirtytalk}
\usepackage{siunitx}
\usepackage{l3draw}

\usepackage{expl3}
\usepackage{tikz}
\usepackage{xcolor}
\usepackage{xparse}
\usepackage{siunitx}
\usepackage{float}
\usepackage{rotating}
\usepackage{pdflscape}
\usepackage{adjustbox}
\usepackage[colorlinks=true, linkcolor=black, citecolor=black, urlcolor=black]{hyperref}

\NewDocumentCommand{\priority}{O{black} m}{%
  \begin{tikzpicture}[scale=0.12]
    \draw (0,0) circle (1);
    \fill[fill opacity=1,fill=#1] (0,0) -- (90:{#2>0?1:0}) arc (90:90-#2*3.6:1) -- cycle;
  \end{tikzpicture}%
}

\definecolor{myyellow}{RGB}{224, 198, 52}
\definecolor{mygreen}{RGB}{0, 105, 98}
\definecolor{mybrown}{RGB}{77, 46, 28}
\definecolor{myorange}{RGB}{230, 122, 7}
\definecolor{myblue}{RGB}{65, 114, 159}
\definecolor{mygray}{RGB}{143, 146, 148}
\smartqed  
\usepackage{graphicx}
\reversemarginpar
\begin{document}

\title{Open Source Software Development Tool Installation}
\subtitle{Challenges and Strategies For Novice Developers}


\author{Larissa Salerno         \and
        Christoph Treude        \and
        Patanamon Thongtanunam
}


\institute{Larissa Salerno \at
              The University of Melbourne \\
              Melbourne, Australia \\ 
        \email{lsalernodeca@student.unimelb.edu.au}
           \and
           Christoph Treude \at
              Singapore Management University \\
              Singapore, Singapore \\
              \email{ctreude@smu.edu.sg}
           \and
           Patanamon Thongtatunam \at
            The University of Melbourne \\
            Melbourne, Australia \\
            \email{patanamon.t@unimelb.edu.au}             
}

\date{Received: date / Accepted: date}

\maketitle

\begin{abstract}
As the world of technology advances, so do the tools that software developers use to create new programs. In recent years, software development tools have become more popular, allowing developers to work more efficiently and produce higher-quality software. Still, installing such tools can be challenging for novice developers at the early stage of their careers, as they may face issues such as compatibility problems (e.g., with operating systems) and unclear instructions. Therefore, this work aims to investigate the challenges novice developers face when installing software development tools and the strategies they employ to overcome them. To investigate these, we conducted an analysis of 24 live software installation sessions to observe the difficulties developers encounter, the strategies they apply, and the types of information sources they consult when facing obstacles. We also conducted a validation survey with 144 students to support and expand our findings. Our results reveal recurring challenges such as unclear or dysfunctional documentation, complicated installation processes, version incompatibility, and lack of feedback during installation. To address these, participants used strategies such as reformulating search queries, reading instructions more carefully, and searching for alternative sources of documentation. These sources included community platforms  (e.g., Stack Overflow), video tutorials, blog posts, and official documentation. Based on these findings, we provide practical recommendations for tool vendors, tool users, and researchers to improve the installation experience for novice developers.


\subclass{MSC code1 \and MSC code2 \and more}
\end{abstract}

\section{Introduction}
\label{intro}

Open Source Software (OSS) are systems with freely accessible and modifiable source code, allowing unrestricted redistribution, derivative works, and use in any field. Licenses must ensure non-discrimination, technology neutrality, and compatibility with other software, without imposing royalties or product-specific conditions \citep{opensource}. A software development tool is any software application program that a user invokes to perform a task associated with a software development project \citep{riddle2012software}. Software development tools play an essential role in software development processes. They help software teams in the development of features, as well as in other activities such as monitoring the progress of software tasks \citep{ramirez2020descriptions}. These tools support developers in many different ways; for instance, Astah UML helps software developers with architecture design, and Git supports versioning of source code. Additionally, each tool has its own features and sometimes different purposes, which may be better suited for specific contexts or projects. Installing software development tools can be considered an intricate process \citep{hassan2017mining}, given that they may have specific system requirements (e.g., compatible machine environments), various forms of installers (e.g., a binary file, source code), and numerous steps of installation. Therefore, developers may encounter issues during the installation process.

A novice developer is an individual who is learning programming for the first time. This population shares common characteristics, such as the inability to formulate proper algorithms to solve given tasks and challenges in mastering the syntax and semantics of programming languages \citep{mselle2012impact}. Recent studies have shed light on how installing software development tools is a challenge for novice developers. For example, research conducted by Steinmacher \citep{paper1} pointed out that one of the barriers newcomers face when first contributing to OSS projects is the hurdles of configuring the development environment, which also involves the installation of tools. Other studies \citep{valez2020student,jenkins2012perspectives} also highlight the hurdles that junior developers face when installing tools. In both papers, participants complained that they sometimes spent hours or even days installing and configuring software development tools. However, little work has explored the challenges of tool installation and the strategies used to overcome them.

In this work, we conduct qualitative research to investigate the challenges novice developers face when installing software development tools and comprehend the decisions and actions they make when encountering such problems. More specifically, we conduct think-aloud sessions to observe participants while they install the tools. We analysed the data using qualitative analysis and extracted from the transcripts the challenges and strategies from each session. Additionally, we conducted a survey to validate our findings. The survey focused on the respondents' past experience with the challenges and strategies identified in the analysis of the think-aloud sessions.






To guide our qualitative analysis, we formulated research questions focusing on three main aspects: 1) Challenges, 2) Strategies, and 3) Source of Information. Based on 24 think-aloud sessions with 18 participants across seven tools, along with a validation survey involving 144 respondents, we address the research questions as follow: \textbf{RQ1} \textit{What are the challenges during the installation process of software development tools?} The common challenges participants faced were related to the unavailability of simpler ways to install the tools, and also the lack of installation progress feedback, meaning they were unsure if the installation was completed successfully or not. Our validation survey also shows that the majority of participants reported finding unclear instructions when installing software development tools, as well as poorly presented information. \textbf{RQ2} \textit{What strategies do participants apply to overcome the challenges?} In terms of strategies, searching error messages or solutions for problems on the web and seeking help or feedback in the terminal were one of the most recurring strategies. Participants in our survey also reported seeking help through the terminal command-line interface as a common strategy. \textbf{RQ3} \textit{What sources do participants consult when they face such challenges?} Regarding the sources, seeking help from online community forums (e.g., Stack Overflow and GitHub Community) proved to help the participants overcome some of the challenges. 

\textbf{Novelty \& Contributions.} To the best of our knowledge, our research is the first to specifically investigate the challenges novice developers face during this process. Additionally, our study explores the strategies these developers find helpful and the types of information they seek when installing software development tools. Identifying the challenges novice software developers face when installing tools could help software organisations better understand how the installation process of software development tools can be enhanced, making the installation process less difficult for both novice and inexperienced software developers. The findings can also help future work in creating potential solutions to these problems, such as establishing a tool or a guideline, therefore speeding up the software development process and increasing tool accessibility and quality. Additionally, they could contribute to creating techniques that automatically extract reliable instructions and present solutions to common installation errors encountered by users.

To address challenges and strategies, we provide a list of recommendations for tool vendors, tool users, and researchers. For tool vendors, we recommend incorporating progress feedback throughout the installation process, since our results showed that this is a significant challenge for developers. For tool users, we suggest consulting alternative sources other than the official documentation, as it usually lacks guidance on recovering from error scenarios. For researchers, future work can explore automated techniques for tool vendors, focusing on eliminating duplicated instructions, providing automatic progress feedback, and automatically testing the installation instructions. 


\section{Related Work}
\label{sec:relatedwork}

The installation of software development tools can be a challenging task for developers, especially for newcomers in the field, as they may face obstacles along the process (e.g., complex tool configuration and documentation). To the best of our knowledge, none of the related work has investigated the actual challenges software developers face when they install software development tools, as well as the strategies they apply to overcome challenges. In this section, we present the findings of the related work that motivates our study. We review research related to Newcomers and Open Source Contributions since installation challenges have often been considered as one of the many problems newcomers encounter when trying to join OSS projects, although none of the related work has focused on the actual installation challenges. Work related to Web Integrated Development Environments (Web IDEs) is relevant, which might make installation processes obsolete, and Documentation is one of the key sources that novice developers consider when installing software development tools.

\subsection{Newcomers and Open Source Contributions}


Challenges in installing software development tools are still not well understood in the industry, though some studies have explored similar issues in the context of OSS projects. Recent studies have shed light on how installing software development tools is challenging for developers. Through a literature review, researchers identified 20 studies that provide empirical evidence of barriers newcomers often face when making a contribution \citep{paper1}. Barriers such as the set-up of the local environment, particularly the complexities associated with installing additional software development tools have shown to be a significant barrier for newcomers seeking to contribute effectively to OSS projects. A follow-up study \citep{paper2} explored the effectiveness of FLOSSCoach, a portal designed to address challenges identified in the previous research. The findings indicated that while the portal successfully mitigated some social challenges, particularly communication barriers, it was less effective in overcoming technical issues such as workspace setup and documentation. The lack of built-in tools or mechanisms to support newcomers with these technical hurdles limited its impact. Participants reported experiencing “frustration, irritation, or demotivation” when encountering these challenges.

From the results of the two studies mentioned previously, the researchers developed a guideline \citep{steinmacher2018let} for communities seeking to support newcomers and also for newcomers who want to start contributing to OSS projects. Regarding the hurdles of the local environment setup, the researchers recommend OSS communities make it easy for newcomers to build the system locally. They emphasise that setting up the local workspace was the most reported challenge in their first study, and this barrier reinforced how newcomers feel frustrated and demotivated, even after using the FLOSSCoach portal. Researchers suggest creating detailed step-by-step tutorials that include links to information about common problems and potential solutions, such as a FAQ section.

Another study collected students’ perceptions regarding the use of OSS projects in the teaching process of software engineering courses \citep{pinto2019training}. The researchers conducted semi-structured interviews with 21 students (undergraduate and graduate level) from five different universities. Despite the students' positive impressions, the findings revealed that setting up the development environment remained a significant challenge for many participants. One participant described it as ``not uncommon to have a headache to set up the project''. Similar challenges were reported in a study that investigated the barriers students faced when taking a course on OSS development \citep{salerno2023barriers}. This indicates that the process of setting up the environment can demotivate students to contribute to OSS projects. Students often struggle to gain industry-focused, hands-on experience in software development before entering the workforce \citep{toth2006experiences}. To address this, a proposed pedagogy aims to enhance the practical aspects of software engineering education by identifying ways educators can effectively integrate OSS tools and components. One finding highlights that OSS projects rarely provide the necessary guidance and resources for students to engage with them.

Prior research indicates that software development tool installation is a significant hurdle for newcomers, with challenges ranging from insufficient documentation and instructions to dependency management issues \citep{steinmacher2015systematic, stol2010identifying, pinto2019training, paper1, toth2006experiences}. However, no studies have documented the specific challenges they face during installation or the strategies they use to overcome them, which are gaps that our study aims to address. While effective tools have been developed to mitigate social barriers \citep{paper2, steinmacher2018let}, there is a limited understanding of strategies that help developers navigate technical challenges during tool installation.

\subsection{Web Integrated Development Environments}

Although the previous studies mentioned above are strictly about OSS projects, a recent study \citep{valez2020student} presents KODETHON, a Web IDE application that was developed to help students concentrate on learning programming without the hassle of installing and configuring tools \citep{valez2020student}. The application has been used by more than 3,000 students from 15 different courses. According to the findings, 61\% of participants reported that one of the valuable features of the application is not having to install and configure anything on their machine. Similarly, JavaWIDE, another web IDE, was created to support programming education \citep{jenkins2012perspectives}. Both students and teachers highlighted how having a tool that does not require installation or configuration eliminated a considerable number of technical problems that they could have faced.

Similarly, Online Python Tutor was developed as a web-based tool that requires no software or plugin installation \citep{guo2013online}. The tool has gained popularity, with over 200,000 people, including instructors and students, using it for teaching and learning purposes. According to students, this tool was seen as a major benefit, as it eliminates the need for time-consuming installations and configurations, allowing students to focus on learning and programming tasks instead. In parallel, a study \citep{helminen2013recording} reported challenges students face during the process of learning programming, and they share a similar opinion that Web IDEs can be helpful.

The studies discussed in this subsection highlight how web-based development environments can support students in learning programming by eliminating the need for tool installation and local environment configuration \citep{valez2020student, jenkins2012perspectives, guo2013online, helminen2013recording}. Although Web IDEs can be a solution to potential challenges during tool installation, these previous studies do not conduct an investigation of the actual challenges faced by novice developers. In addition, it is important to note that Web IDEs are not a remedy for all software development tool installation scenarios. Not all software packages and development tools can be seamlessly integrated or utilised within a Web IDE framework due to limitations in accessing hardware resources, complexities in simulating specific development environments, or restrictions in customising the development setup to match real-world conditions. This motivates us to better understand the challenges involved in the software development tool installation process.

\subsection{Documentation}

Researchers have been investigating the issues that practitioners in Software Engineering (SE) encounter in software documentation \citep{uddin2015api, chen2009empirical}. In a first study \citep{aghajani2019software}, the researchers mined the problems related to documentation in four sources: Issues, mailing lists, pull requests, and Stack Overflow. After manually labelling the data, they identified 162 software document issues. In order to validate the results from the study, the researchers conducted a second study \citep{aghajani2020software} where they validated the 162 taxonomies with 146 professional software practitioners. One of the most significant results was regarding \underline{\textit{installation instructions}}, where 65\% of the participants reported that \textit{faulty tutorial} is a problem they usually find in software documentation. A second aspect is the \textit{inappropriate installation instructions}, which was the most frequent issue according to the practitioners. This shows that software documentation is a problem for developers when they are installing software development tools. Another study \citep{gao2025add} investigated installation instruction updates along with the triggering behind them in a rapidly evolving software ecosystem. The researchers further propose the automated updates of README files to incorporate installation steps, which was found to be useful by both documentation maintainers and downstream users.  A similar study \citep{gao2023evaluating} has focused on simplifying README files to make them more comprehensive for software developers.

Studies report through the perspective of software engineering practitioners challenges in software documentation \citep{uddin2015api, chen2009empirical}. One study \citep{aghajani2019software} identified 162 documentation issues by analysing four sources: issue trackers, mailing lists, pull requests, and Stack Overflow. A follow-up study \citep{aghajani2020software} validated these findings with 146 professional software practitioners, highlighting installation instructions as a major concern. Among the participants, 65\% reported encountering faulty tutorials, while inappropriate installation instructions were identified as the most frequent issue. These findings reinforce the impact of documentation quality on developers' ability to install software development tools. Despite the availability of guidelines and automated tools to extract information from software \citep{liu2022readme, hassan2017mining, robillard2017demand}, software developers often do not create clear and reliable documentation. One of the contributions of our study is to lay the groundwork for a technique that can automatically extract reliable instructions and also present solutions to the most common errors that users encounter when installing software development tools.

To assess the executability of software tutorials, a study analysed over 600 step-by-step tutorials, examining whether they could be followed to completion. The results showed that with a naive execution strategy, only 26\% of tutorials were executable, and even with a more sophisticated approach using human annotation, the executability rate only increased to 52.3\%. Identified issues included inaccessible resources, missing instructions, inconsistent file handling, and poor documentation quality \citep{mirhosseini2020docable}. A study on Application Programming Interfaces (APIs) highlights the dispersion of documentation across the internet \citep{treude2018does}, which can be confusing for developers who are seeking software documentation. To address this issue, the researchers replicated a previous work on the topic \citep{parnin2011measuring} and complemented the work by selecting ten popular APIs with the aim of covering a variety of programming languages such as Java, Javascript, PHP, C++, and Python. The study found that, in addition to official documentation, GitHub and Stack Overflow are the sources that appear most frequently on the first page of Google search results for the selected APIs. This indicates that developers often rely on these platforms for finding information related to APIs. However, the dispersed documentation across multiple sources can also lead to duplicate information, which can further confuse developers.

Previous studies indicate that creating and maintaining step-by-step tutorials is difficult \citep{lafreniere2013community, krosnick2015videodoc, pavel2014video}, and in an attempt to overcome some of the challenges, tools have been developed \citep{mysore2017torta, lafreniere2013community}. The studies mentioned in this section highlight several documentation issues and also introduce tools designed to overcome some of these problems. To better understand how novice developers search for information in software development tool installation, we analyse and map the types of sources they consulted during think-aloud sessions, connecting them with the challenges they encounter and the strategies they apply.

\section{Methodology}
\label{sec:methodology}

We used qualitative research methods (i.e., surveys and observation) to conduct the study, as well as a number of tools to support data collection and analysis. To guide our research, we addressed the following research questions:

\begin{itemize}
\item \textbf{RQ1. \textit{What are the challenges during the installation process of software development tools?}} So far, no studies have focused on explicitly identifying the hurdles novice developers face when installing such tools. We aim to identify the challenges via software installation sessions, where participants will be asked to install a software development tool.

 \item \textbf{RQ2.\textit{ What strategies do participants apply to overcome the challenges?}} Understanding the strategies they employ, we can identify effective or non-effective strategies that can support other software developers and tool vendors.

 \item \textbf{RQ3. \textit{What sources do participants consult when they face such challenges?}} Our objective is to identify sources on which participants rely when implementing strategies to overcome the challenges. As part of this, we study the relationship between challenges, strategies, sources, and tools. 
  \end{itemize} 

In this section, we describe the participants of our think-aloud study (Subsection \ref{sec:participants}), the data collection (Subsection \ref{sec:datacollection}) and analysis procedures (Subsection \ref{sec:datanalysis}), as well as the methodology of our validation survey (Subsection \ref{sec:validationsurvey}).
  
\subsection{Participants}
\label{sec:participants}
After obtaining approval from our institution's ethical review board, the participants for our study were university students undertaking a computer-related degree. We have decided to choose university students close to graduation, as previous work, such as the study carried out by \citet{paper4}, evinces that students can perform the role of professional developers in software engineering experiments. The challenges encountered by students close to graduation are a good proxy for the challenges encountered by professionals at the beginning of their careers. The recruitment process occurred through posting announcements about the research and verbally advertising the research in computing and/or software engineering classes. To express interest in participation, students submitted a demographic survey, and qualified students (e.g., undertaking a computer-related degree) were invited to participate. To incentivise participation, participants were offered compensation in the form of a gift card to attend the think-aloud sessions to install software tools.

Researchers have provided empirical evidence that students can, under specific contexts, be a suitable sample for conducting experiments in software engineering \citep{falessi2018empirical}. In one study, 64 experts in software engineering experimentation participated in focus group sessions where they reflected on their experiences conducting experiments with both professionals and students.
The findings showed that, while students are not universally representative of professionals, they can serve as appropriate proxies in specific contexts, such as in early-stage experiments, tasks that do not require deep domain knowledge, or when students have relevant industrial experience. Another study also identified contexts in which students may represent professionals, such as in typical start-up environments \citep{fagerholm2013platform}. This view is further supported by another study, which reported that in two of their experiments, 40\% of student participants were either employed part-time or had previous full-time work experience, reinforcing the notion that students can be representative of professionals \citep{sjoberg2002conducting}.

In total, 18 novice developers participated in the study, of which 13 were Master's students and the other five were undergraduate students who were majoring in a computer-related discipline. We consider this sample size to be satisfactory given the time and effort required to conduct and analyse the think-aloud sessions. With this group of participants, we were able to run the study with participants with different levels of knowledge and experience (see Table \ref{tab:table1}). The number of participants is also similar to previous qualitative studies \citep{thiselton2019enhancing, robillard2020understanding} that conducted think-aloud sessions with 16 and 18 participants.

\subsection{Data Collection}
\label{sec:datacollection}
\begin{figure}[ht]
\centering
\includegraphics[width=\textwidth]{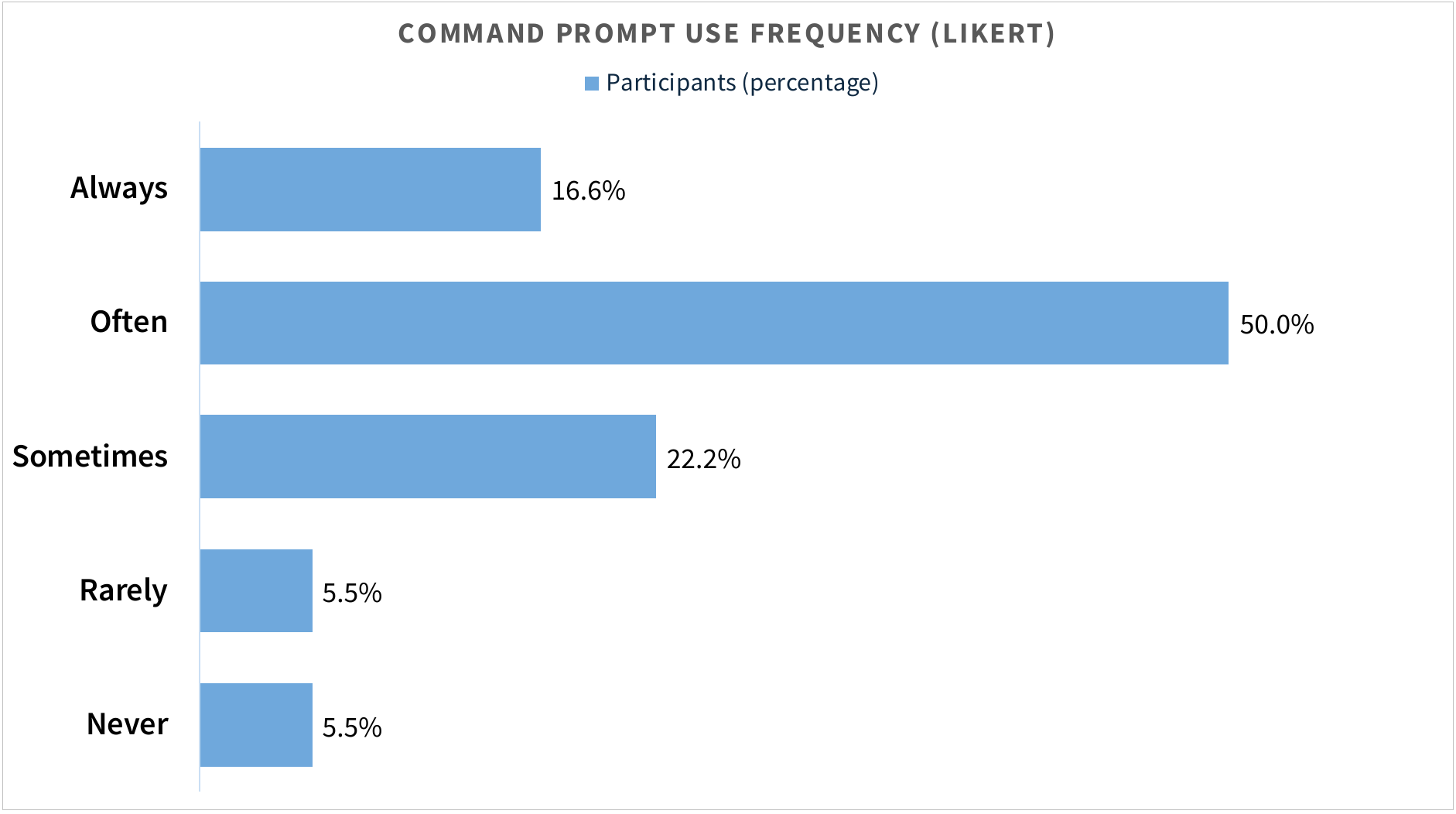}
\caption{Results for the survey question ``Tell us how frequently you use the command prompt for installing software development tools?''.}
\label{fig:frequency}
\end{figure}

\begin{figure}[ht]
\centering
\includegraphics[width=\textwidth]{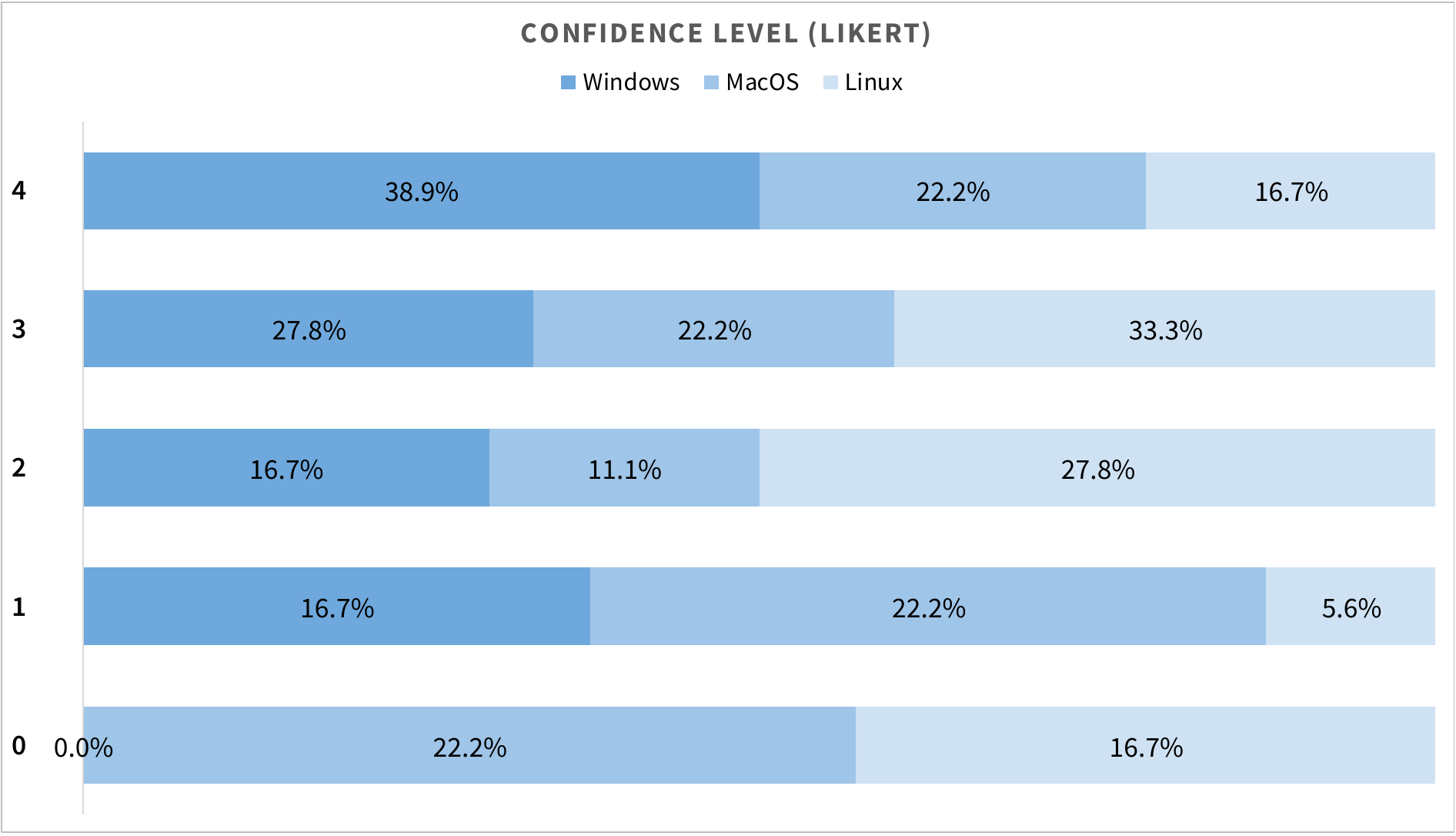}
\caption{Results for the survey question ``How confident are you in installing software development tools using the operating systems below?''.}
\label{fig:confidence}
\end{figure}

Each participant individually responded to a short questionnaire and then participated in live software installation sessions in a think-aloud format. As shown in Table \ref{tab:table1}, Figure \ref{fig:frequency} and Figure \ref{fig:confidence}, the short questionnaire included questions concerning their degree, their previous experience installing software development tools using different operating systems (e.g., Windows or Linux), and the tools they would like to install during the think-aloud session. A list of seven software development tools was provided in the questionnaire. 
Participants could choose any tool in the list.
These tools were selected by the authors based on their popularity on the OSSRank website \citep{ossrank}, which highlights widely used OSS tools among developers. We also considered the relevance and potential benefit of each tool to the participants' coursework. 
During the sessions, we ensured that each tool was installed by at least three different participants. This approach aimed to balance participant interest with experimental consistency, ensuring the task was meaningful while maintaining some control.

The categories of tools chosen for this study were Python libraries, Frameworks, and Database. For the Python libraries, we selected TensorFlow, PyTorch, and Jupyter Notebook. These tools are commonly used in machine learning and deep learning contexts. \textbf{TensorFlow}\footnote{\cite{tensorflow}} and \textbf{PyTorch}\footnote{\cite{pytorch}} are both frameworks designed for tasks such as predictive modelling and natural language processing. \textbf{Jupyter Notebook}\footnote{\cite{jupyter}} is an interactive environment that allows users to create and share documents containing live code, equations, visualisations, and narrative text. All three tools require Python to be installed, and they come with additional dependencies that are automatically installed depending on the installation method (e.g., pip or conda).

For the frameworks, we selected Flutter, Node.js and Spring Boot. \textbf{Flutter}\footnote{\cite{flutter}} is a UI (User Interface) toolkit for building natively compiled applications for mobile, web, and desktop systems from a single codebase, using the Dart programming language. \textbf{Spring Boot}\footnote{\cite{springboot}} is a Java-based framework designed to simplify the creation of enterprise applications and microservices. \textbf{Node.js}\footnote{\cite{nodejs}} allows developers to run JavaScript on the server side, making it ideal for building scalable network applications. Each of these tools requires its respective programming environments; Dart for Flutter, Java for Spring Boot, and JavaScript for Node.js and can also require additional dependencies which are installed automatically with the tool.

\textbf{MongoDB}\footnote{\cite{mongodb}} is a NoSQL database used to handle large volumes of unstructured data. MongoDB requires the installation of its own server, as well as familiarity with JavaScript. In addition to the core database, MongoDB may also require additional tools to be installed (MongoDB Compass\footnote{\cite{compass}}, MongoDB Shell\footnote{\cite{mongosh}}), which are also typically installed automatically during the setup process.

\begin{table}[ht]
\centering
\caption{Demographics of participants.}
\label{tab:table1}
\resizebox{\textwidth}{!}{
\begin{tabular}{@{}lll@{}}
\toprule
\textbf{Survey Questions} & \textbf{Type} & \textbf{Distribution of Answers} \\ \midrule
What degree are you currently studying? & Multiple choice & \begin{tabular}[t]{@{}l@{}}Master of Information Technology: 33.3\%\\ Master of Software Engineering: 33.3\%\\ Bachelor of Science: 27.7\% \\ Master of Computer Science: 5.5\%\end{tabular} \\ \midrule
\begin{tabular}[t]{@{}l@{}}What operating system do you prefer for using \\ software development tools?\end{tabular} & Multiple choice & \begin{tabular}[t]{@{}l@{}}Windows: 44.4\%\\ MacOS: 44.4\%\\ Linux: 11.1\%\end{tabular} \\ \bottomrule
\end{tabular}}
\end{table}

Inspired by the study of~\citet{paper3}, we conducted live software installation using a think-aloud format. We consider the think-aloud format a suitable method for this study, since we can observe the actual behaviours of the participants along with their thoughts. We scheduled individual appointments with each participant (referred to as \textit{participant appointment(s)}), during which we asked them to install a software development tool of their choice from the predefined list in the questionnaire. The individual tool sessions are being referred in the paper as \textit{installation sessions}, \textit{think-aloud sessions}, or \textit{sessions}.

At the beginning of each participant appointment, we explained the concept of a think-aloud session and asked participants to verbalise their thoughts throughout the installation process. Supported by previous research \citep{chattopadhyay2018context}, each participant appointment was limited to one hour to prevent fatigue, which could compromise the quality of data collection. If a participant finished the installation before the end of the scheduled time, they were offered the opportunity to install an additional tool. To ensure they had enough time to reasonably engage with the second installation, we only offered a second installation if the first installation finished within approximately 20 to 25 minutes. Most of the second installation were completed within the 1-hour session. In one case (P8), the session had to be concluded at the one-hour mark due to scheduling constraints, even though the participant was still attempting the installation of the first (and only) tool. From an ethical standpoint, participants were monetarily compensated with an amount that fairly reflected one hour of their time. Additionally, novice developers often need to install software development tools when starting a new job, which is why we believe this task reflects a realistic setting. 


Participants were asked to bring their own machine to the session, as this would reflect a potential settings in practice where developers often work with their personal or already-configured devices. However, we acknowledge that in some professional contexts, receiving a clean machine is also a realistic scenario. The operating systems used, as shown in Table \ref{tab:table1}, were Windows, Linux, and macOS. Since participants were using their own machines, we assumed they had at least a basic understanding of the installation procedures on their respective devices. When installing the tool, participants were allowed to use the method of their choice and consult any sources they preferred for installation instructions. A researcher observed each participant’s progress throughout the session, taking notes on their actions and thoughts. At the end of the session, the researcher asked what challenges they faced during the installation. While participant thoughts were important, our primary focus was to capture real-time behaviour. For this reason, the observer frequently asked prompting questions such as \textit{“What is happening now?”} or \textit{“Have you ever faced this (challenge) before?”}.

In order to support data collection, several tools were used to record actions and interviews during the observation sessions. We conducted the sessions in soundproof meeting rooms equipped with a large table, chairs, and the necessary recording equipment. The actions related to software installations were observed and captured via screen recording software to track their actions during the installation tasks. We also made use of audio recordings to keep track of participants' think-aloud thoughts. Screen and voice recordings were captured using Zoom. As a backup plan, we video-recorded the sessions to help researchers during the analysis process of the think-aloud sessions (e.g., if the participant was pointing at the screen to show a determined aspect). The camera was placed in a way that did not record the participant's face.

\subsection{Data Analysis}
\label{sec:datanalysis}

\begin{figure}[ht]
\centering
\includegraphics[scale=0.44,page=1]{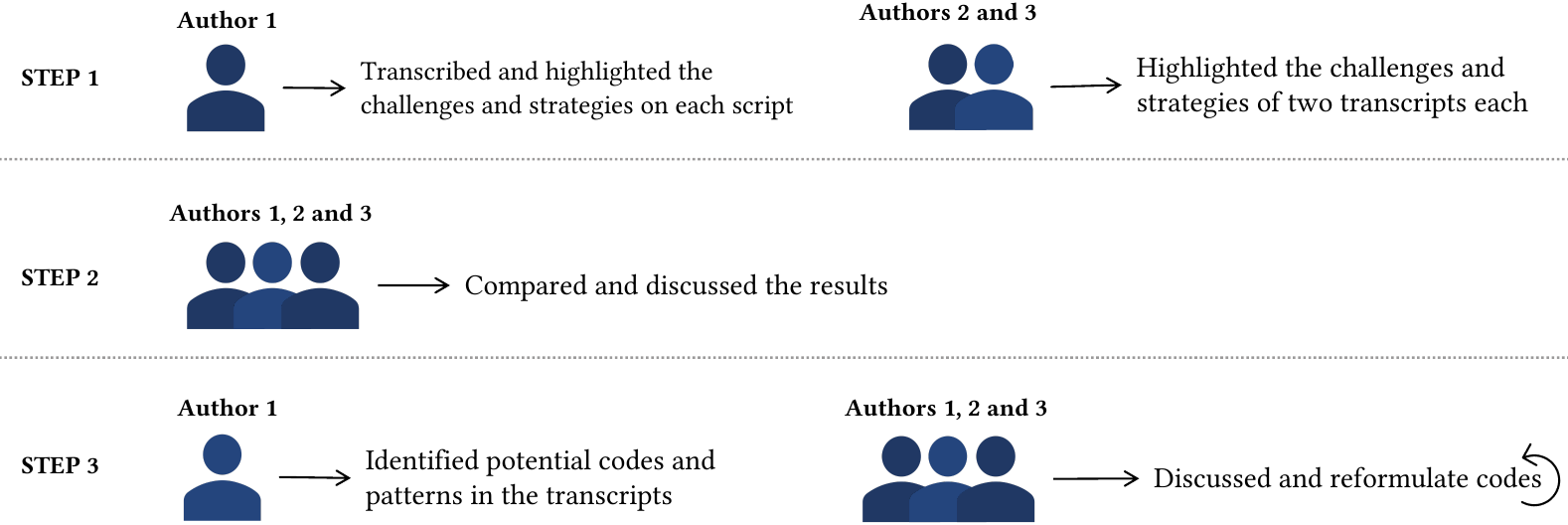}
\caption{Data analysis process.}
\label{fig:dataanalysis}
\end{figure}

We used qualitative analysis to analyse the data collected during the installation sessions. In total, we conducted 24 installation sessions, in which we collected \textbf{534 minutes} of video recording and \textbf{187 pages} of transcripts. The analysis process was divided into 3 steps, as shown in \textbf{Figure \ref{fig:dataanalysis}}. In step 1, Author 1 transcribed the think-aloud sessions. Then, Author 1 identified and highlighted the challenges and strategies reported by the participants in all transcripts. In step 2, to reduce potential bias from having a single annotator, Author 2 and Author 3 independently identified and highlighted the challenges and strategies in two of the same transcripts. They then compared their annotations with those of Author 1 and had a meeting to resolve any disagreements. After achieving consistency in the second transcript, where only minor differences were found and addressed, Author 1 proceeded with coding the remaining transcripts. Throughout this process, Authors 2 and 3 regularly reviewed and validated all codes annotated by Author 1. 

In step 3, Author 1 started extracting codes from the challenges and strategies highlighted in the transcripts. Then, all authors discussed and reformulated the codes from all sessions iteratively. Author 2 and Author 3 read and analysed the first transcript and resolved potential conflicts. Then, for the second transcript, all 3 annotations were consistent with very few conflicts that were resolved, suggesting that the annotation is now more consistent. Thus, Author 1 continued to do the remaining. Nonetheless, Authors 2 and 3 reviewed all the codes annotated by Author 1 iteratively.

In the annotation process, we first identified the sentences or phrases that could be classified as a challenge or strategy, referring to the video recordings. Each identified challenge and strategy was assigned a name. Once this process was completed for all transcripts, we carefully merged similar codes and refined them iteratively. In total, we extracted \textbf{23} codes, where \textbf{10} are related to the \textbf{challenges} developers face when installing software development tools, and \textbf{13} are the \textbf{strategies} they employ when faced with these challenges.

\begin{figure}[ht]
    \centering
    \includegraphics[scale=1.3]{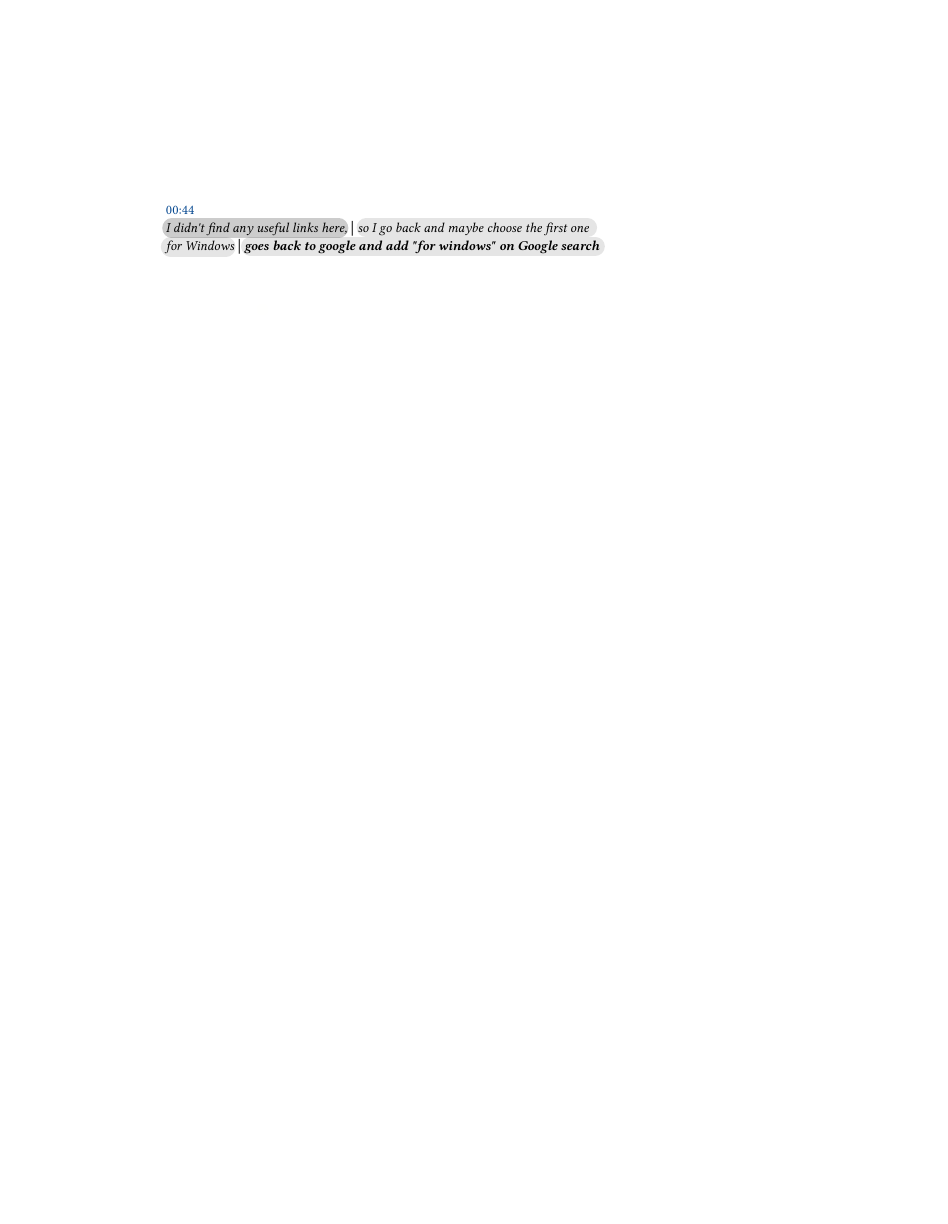}
    \caption{Example of code extraction.}
    \label{fig:transcript}
\end{figure}

Figure \ref{fig:transcript} illustrates an example of our code extraction process for an excerpt taken from one of the transcripts.  The text highlighted in dark grey is identified as a challenge, the text highlighted in light grey is identified as a strategy, and the text in bold represents the participant's action. In the given example, a participant was searching for installation instructions on Google. Initially, the participant phrased the query by specifying the tool name along with the word 'download'. After clicking on one of the links suggested by Google, the participant found the instructions quite generic and decided to return to the Google search, reformulating the query by adding the operating system to the search terms. As a result, we identified the challenge \textit{``I didn't find any useful links here,|''} as part of the code ``Under-specified Query'', and the strategy \textit{``so I go back and maybe choose the first one for Windows | \textbf{goes back to Google and adds ``for windows'' on Google search}''} is related to the code ``Reformulating the Query''. Towards the end of the analysis process, we realised there was a saturation of codes, as no new major codes were identified with the last few participants. As the coding process progressed, we organised the challenges and strategies into categories that emerged from the coding process. To support transparency and enable further exploration, all interview transcripts, coding results, and validation study outcomes are available in the publicly accessible repository (see Section \ref{sec:declarations}).

\subsection{Validation Survey}
\label{sec:validationsurvey}
To validate our findings from the think-aloud sessions, we conducted a survey with a larger group of students to substantiate the findings of our study. The questionnaire was distributed to 148 third-year undergraduate university students who took a software engineering course with a focus on software processes. In this course, students were introduced to OSS development, including the history of OSS development, processes, practices, licenses, and tools. These topics were addressed through lectures and practical exercises, covering aspects such as source code management, code review, and continuous integration using GitHub.

In the questionnaire, we listed the challenges and strategies that we identified from the think-aloud questions and asked them to indicate the challenges they had previously encountered and the strategies they had applied to address the challenges. The challenges and strategies were presented in two independent questions to avoid potential bias. We phrased the survey question as: ``Please select the challenges you have faced when installing software development tools in the past (Choose all that apply)'', which included a multiple-selection list of all the challenges identified in the study. Following this, the second question asked: ``Please select the strategies you usually apply when you face challenges installing software development tools (Choose all that apply)'', which provided a multiple-selection list of all the strategies. Both challenges and strategies are presented in descending order based on the number of think-aloud sessions in which participants encountered them (as shown in Tables \ref{tab:challenges} and \ref{tab:strategies}),
along with their respective descriptions. Additionally, the survey questions, along with the descriptions of challenges and strategies as presented to the survey participants, are available in our replication package (see Section~\ref{sec:declarations}). Note that the challenge and strategy categories were not visible to participants in the questionnaire. For example, the survey showed the challenges ``Unclear Instructions'' and ``Dysfunctional Instructions'', but not the corresponding category ``Information Quality/Value'', since we did not want to bias participants by exposing details of our qualitative analysis process. As part of the questionnaire, we offered participants ``Other(s)'' and ``None of the above'' options for each question, just in case they wished to include additional challenges or strategies. We do not include these additional challenges in Table 3, as each was reported by only a single participant.

In total, we received 144 responses, from which we analysed and counted how many times a challenge and strategy were selected. The results of the survey can be found in Section 4 (see Table \ref{tab:challenges} and Table \ref{tab:strategies}). The results are displayed in the tables listing the \textbf{Challenges} and \textbf{Strategies} as a column labelled ``Survey''. In both tables, a few of the challenges and strategies are marked with ``-'', indicating their absence in the survey as they emerged from the think-aloud sessions with the four participants whose sessions we scheduled after we sent out the survey. This occurred because we sent the survey after conducting 14 participant appointments. After analysing the results, we decided to recruit four more participants for additional think-aloud sessions to balance the installation of Spring Boot, Jupyter Notebook, Node.js, MongoDB, and Flutter, as we wanted to have at least three participants per tool. As a result, the survey participants did not see the challenges and strategies that emerged from these later sessions. The challenge not displayed to the survey participants was ``Error Message'', and the strategies not displayed were ``Waiting'', ``Random Actions'' and ``Ignoring''.

\section{Results}
\label{sec:results}

In this section, we present our findings and provide details from the think-aloud study to illustrate specific challenges and strategies. In our study, 18 Master's and undergraduate students participated in software installation sessions in a think-aloud format. In total, seven software development tools (listed in Table \ref{tab:table2}) were installed. Thirteen participants installed one tool, four participants installed two tools, and one participant installed three tools in a session, resulting in a total of 24 installation sessions. We based our tool selection criteria on the ranking of OSSRank \citep{ossrank} which lists the top OSS development tools within the most popular categories of tools, which included: Python libraries, Frameworks, and Database.



\begin{table}[ht!]
\centering
\caption{List of tools selected for our study.}
\label{tab:table2}
\resizebox{\textwidth}{!}{
\tiny
\begin{tabular}{@{}clllllcll@{}}
\toprule
\multicolumn{1}{l}{\textbf{Tool ID}} &  & \textbf{Tools} &  & \textbf{Type} &  & \multicolumn{1}{l}{\textbf{No. of Installation Sessions}} &  & \textbf{Participants} \\ \midrule
\textit{T} &  & TensorFlow &  & Python Library &  & 4 &  & P1, P6, P7, P11 \\
\textit{F} &  & Flutter &  & Framework &  & 4 &  & P2, P3, P6, P17 \\
\textit{M} &  & MongoDB &  & Database &  & 4 &  & P2, P8, P14, P17 \\
\textit{P} &  & PyTorch &  & Python Library &  & 3 &  & P5, P7, P10 \\
\textit{J} &  & Jupyter Notebook &  & Python Library &  & 3 &  & P4, P13, P16 \\
\textit{N} &  & Node.js &  & Framework &  & 3 &  & P12, P17, P18 \\
\textit{S} &  & Spring Boot &  & Framework &  & 3 &  & P9, P15, P16 \\ \bottomrule
\end{tabular}}
\end{table}



We extracted ten challenges and 13 strategies from the think-aloud sessions listed in Tables \ref{tab:challenges} and \ref{tab:strategies}. In order to simplify the understanding of the results, we organised the challenges and strategies into four categories: Information Quality/Value, Information Seeking, Installation Process, and Tool Characteristics. The category ``Information Quality/Value'' represents challenges and strategies that are related to the installation instructions provided to the user (e.g., instructions from a website). The category ``Information Seeking'' represents information other than official installation instructions, but is still related to the installation that the participants were looking for (e.g., tutorial). The ``Installation Process'' category pertains to the events that occurred when the participants were in the process of or about to initiate the tool installation. Lastly, the ``Tool Characteristics'' category comprises the challenges or strategies that occurred due to a particular attribute of the tool. 

Tables \ref{tab:challenges} and \ref{tab:strategies} provide detailed information as follows: The ``Challenge'' and ``Strategy'' columns specify the names of the challenges and strategies, respectively. The ``Description'' column provides a brief description of each challenge or strategy, while the ``Tools'' column indicates how many participants encountered a challenge or applied a strategy during the installation of each tool. Each dot in the ``Tools'' column corresponds to a specific tool, listed in the following order: TensorFlow, PyTorch, Jupyter Notebook, Flutter, Node.js, Spring Boot, and MongoDB. For ease of reference, this order is also provided in the last row of the table.

To visually represent the number of participants who encountered a specific challenge or used a particular strategy for a given tool, we used a dot system with different colours and fill levels. The fill color indicates how many participants installed a tool: black for four participants and green for three. An empty circle (\priority{0}) indicates zero participants. The fill proportion represents the percentage of participants who encountered the challenge. For instance, a quarter-filled circle (\priority{25}) indicates that one in four participants faced challenges, whereas a third-filled circle (\priority[teal!60]{33}) indicates one in three. 
A fully filled circle (\priority{100} or \priority[teal!60]{100}) represents that all participants installing a tool encountered the challenge. The ``No. of Installation Sessions'' column sums up the total installation sessions in which a challenge or strategy was encountered, and the ``Survey'' column reflects the number of survey participants who reported encountering each challenge or strategy in the past. To visualise the mapping between the challenges and their corresponding strategies, please refer to Table \ref{tab:detailedanalisys}.
\subsection{Challenges (RQ1)}\label{sec:challenge_result}

Table \ref{tab:challenges} lists the ten challenges grouped into four main categories that we identified from our think-aloud sessions.

\begin{table}[]
\centering
\caption{List of challenges the participants encountered during the think-aloud sessions.}
\label{tab:challenges}
\begin{adjustbox}{angle=90,center}
\begin{tabular}{lllcc}
\hline
\textbf{Challenges} & \textbf{Description} & \multicolumn{1}{l}{\textbf{Tools}} & \textbf{No. of Installation Sessions} & \textbf{Survey} \\ \hline
\rowcolor[HTML]{E4E4E4} 
\textbf{Category: Information Quality/Value} & Challenges related to the installation instructions provided to the user &  &  &  \\
\hspace{1em}Unclear Instructions & Installation instructions provided unclear and difficult to follow & \priority{25} \priority{0} \priority[teal!60]{67} \priority{0} \priority[teal!60]{67} \priority[teal!60]{67} \priority{75} & 10 & 88 \\
\hspace{1em}Dysfunctional Instructions & Instructions do not work & \priority{50} \priority[teal!60]{33} \priority[teal!60]{33} \priority{50} \priority[teal!60]{33} \priority[teal!60]{67} \priority{25} & 10 & 62 \\
\rowcolor[HTML]{E4E4E4} 
\textbf{Category: Information Seeking} & Represent information other than official installation instructions &  & \textbf{} &  \\
\hspace{1em}Under-specified Query & Participant searches for information using a non-specific query & \priority{25} \priority[teal!60]{33} \priority[teal!60]{33} \priority{0} \priority[teal!60]{33} \priority{0} \priority{75} & 7 & 33 \\
\hspace{1em}Information Overload & Web page or terminal shows the user too much information & \priority{50} \priority{0} \priority{0} \priority{0} \priority[teal!60]{33} \priority{0} \priority{25} & 4 & 51 \\
\hspace{1em}Poorly-presented Information & Source is hard to navigate and not user-friendly & \priority{0} \priority{0} \priority[teal!60]{33} \priority{0} \priority[teal!60]{33} \priority[teal!60]{100} \priority{50} & 7 & 67 \\
\rowcolor[HTML]{E4E4E4} 
\textbf{Category: Installation Process} & Events that occurred when participants were in the process of or about to initiate installation &  & \textbf{} &  \\
\hspace{1em}Complicated Installation Process & Tool requires multiple steps and/or dependencies to be installed & \priority{50} \priority[teal!60]{33} \priority[teal!60]{33} \priority{75} \priority[teal!60]{33} \priority[teal!60]{100} \priority{75}  & 14 & 52 \\
\hspace{1em}Lack of Installation Progress Feedback & No feedback is provided on the status of the installation & \priority{25} \priority[teal!60]{33} \priority[teal!60]{67} \priority{25} \priority[teal!60]{67} \priority{0} \priority{50} & 9 & 29 \\
\hspace{1em}Error Message & Participant gets stuck with error messages & \priority{50} \priority[teal!60]{33} \priority[teal!60]{33} \priority{0} \priority{0} \priority[teal!60]{67} \priority{50} & 8 & - \\
\rowcolor[HTML]{E4E4E4} 
\textbf{Category: Tool Characteristics} & Comprises the challenges or strategies that occurred due to a particular attribute of the tool & \textbf{} &  & \textbf{} \\
\hspace{1em}Account Required & Tool required the user to create an account on their platform & \priority{0} \priority{0} \priority{0} \priority{0} \priority{0} \priority{0} \priority{50} & 2 & 41 \\
\hspace{1em}Version Incompatibility & CPU or library version is not compatible with the tool & \priority{25} \priority[teal!60]{33} \priority{0} \priority{0} \priority{0} \priority[teal!60]{33} \priority{25}  & 4 & 30 \\ \hline

\multicolumn{5}{l}{The order of the tools is: TensorFlow, PyTorch, Jupyter Notebook, Flutter, Node.js, Spring Boot, and MongoDB} \\ \hline
\end{tabular}
\end{adjustbox}
\end{table}


\subsubsection{Information Quality/Value}

We identified two challenges that fall under the Information Quality/Value category, i.e., Unclear Instructions and Dysfunctional Instructions.
Participants faced the challenge of unclear instructions and dysfunctional instructions in ten sessions each.

\begin{figure}[h!]
\centering
\includegraphics[width=\linewidth]{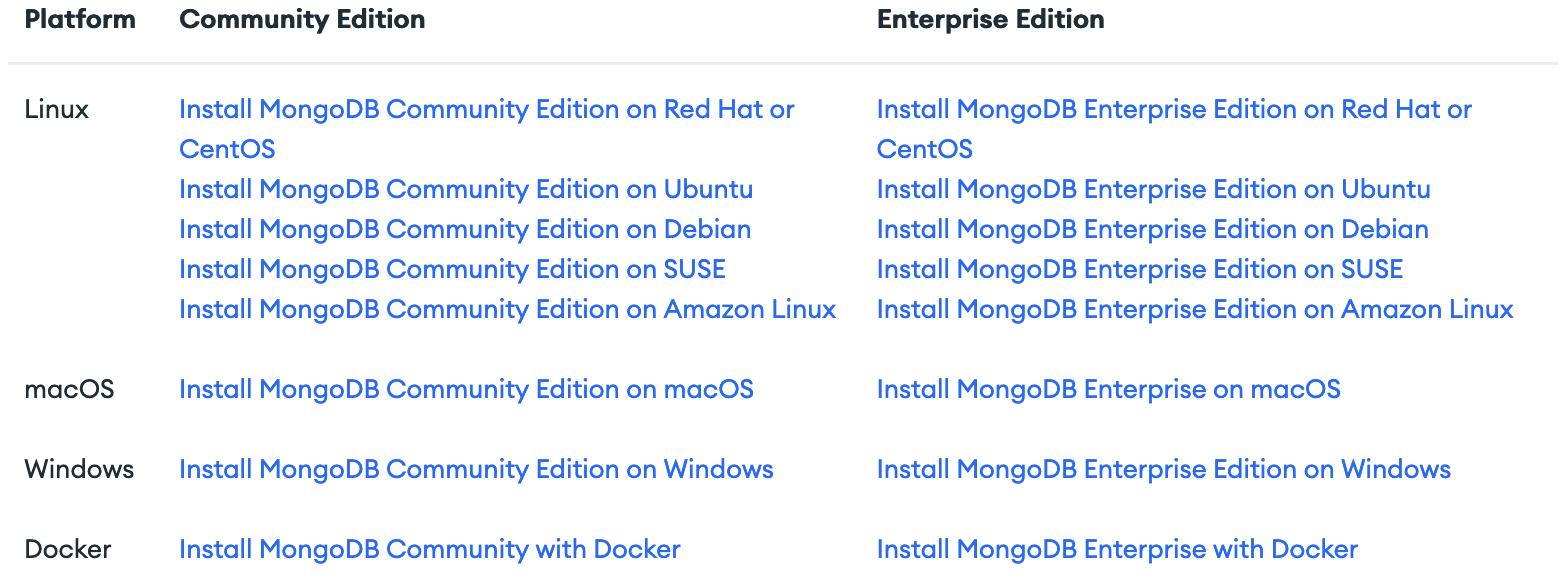}
\caption{Screenshot of MongoDB official website displaying a list of different versions (Accessed on: September 14, 2022).}
\label{fig:mongodbp2-version}
\end{figure}

\underline{\textbf{\textit{Unclear Instructions}}} refers to the challenges when participants read instructions to install the tool, but could not follow or understand due to the somewhat unclear or complicated instructions. This challenge occurred in ten sessions during the installation of TensorFlow, Jupyter Notebook, Node.js, Spring Boot, or MongoDB. In two of the sessions on MongoDB installation, participants searched for instructions to download and install MongoDB using Google search engine. Both participants found that the official website had two different pages containing different instructions on how to install the tool, i.e., one page instructs one to download and install the tool from a zip file and another page instructs to use a package manager (Homebrew) to install the tool.
Such inconsistent and unclear instructions confused the participants: \textit{``Now there is another type of instructions on another page and it says that to install it I need to use Homebrew, and this is annoying'' (P2)}. In another session of installing MongoDB, participants accessed the official website and found the page with the installation instructions. The participants got confused because the page showed two different instances of MongoDB (see Figure \ref{fig:mongodbp2-version}), and the participants did not know which one they were supposed to download $-$\textit{``I don’t know what the difference is [...]'' (P8)}. Three participants faced this uncertainty about the right version when installing MongoDB. It is worth noting that this challenge was faced by two of the three participants (\priority[teal!60]{67}) installing Jupyter Notebook and Node.js, as well as three of the four participants (\priority{75}) installing MongoDB. Besides, 88 respondents in our validation survey reported having encountered unclear instructions when installing a software development tool.

\ul{\textbf{\textit{Dysfunctional Instructions}}} refers to the challenges when the participants followed the installation instructions, but the installation outcomes were not as described in the step-by-step guide. This challenge occurred in ten sessions during the installation of all seven tools, particularly when installing TensorFlow, Flutter, or Spring Boot. When installing TensorFlow, one of the participants followed the installation instructions of the official website as the main source of information. While the participant followed the steps, the participant was required to run a command on the terminal. When the participant ran the command, they got an error message saying that the terminal did not recognise the command. In the MongoDB installation session, the participant followed the instructions on the official website. When the participant copied the command from the instruction page and pasted it into the terminal, they received an error message stating ``Error: Your Command Line Tools are too outdated''. The participant was confused and could not proceed further or solve the problem: \textit{``What? ``Your Command Line Tools are too outdated''? That does not show you any updates on how to do this. Will this take a long time? I do not know'' (P14)}. This challenge happened in two out of three participants (\priority[teal!60]{67}) when installing Spring Boot, and two out of four participants (\priority{50}) when installing Flutter. In our validation survey, 62 respondents revealed having faced this challenge in the past.


\subsubsection{Information Seeking} We identified three challenges that are related to information, i.e., Under-specified query, Information Overload, and Poorly-structured Information.

\underline{\textbf{\textit{Under-specified Query}}} refers to the challenges the participants faced due to unspecified queries when searching for how to install the tool. Participants faced this challenge in seven sessions when installing TensorFlow, PyTorch, Jupyter Notebook, Node.js, or MongoDB. In all of the sessions, the participants tried to search for a web page of installation by querying the search engine Google. However, on the first attempt, they could not find a web page with instructions because they were only searching for the tool name without specific context or purpose, e.g., adding the word ``install'' or ``tutorial''. For instance, when installing TensorFlow, the participant first searched on Google for ``TensorFlow'', and the results did not show any page with instructions. The participant realised that they would need to reformulate the query in order to find the page with instructions $-$\textit{``[..] I'm gonna search ``TensorFlow''. Hum... I'm looking for a Mac version and I want to have a GPU acceleration for it, so I'm gonna search for ``mac GPU TensorFlow.'' (P1)}. We emphasise that this challenge was faced by three out of four participants (\priority{75}) who installed MongoDB. For the other tools, only one participant encountered this challenge per tool. This challenge was also reported by 33 respondents in our validation survey.

\begin{figure}[h!]
\centering
\includegraphics[width=\linewidth]{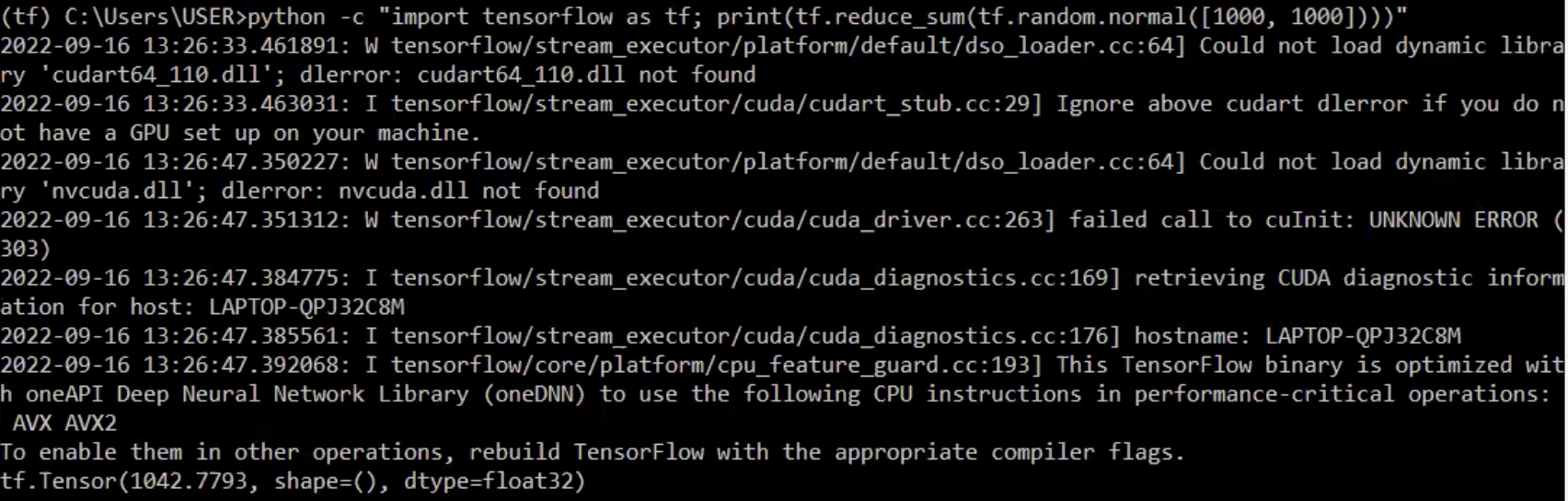}
\caption{Terminal output displaying overloaded information during the installation of TensorFlow.}
\label{fig:tensorflowp11}
\end{figure}

\underline{\textbf{\textit{Information Overload}}} refers to challenges when the web page or a terminal shows excessive information that participants could not follow.  This challenge occurred in four sessions when installing TensorFlow, Node.js, or MongoDB. In one of the sessions, the participant was in the process of installing TensorFlow using the terminal. By running one of the commands according to the instructions, they were surprised with the amount of information from the terminal, and the participant could not understand if the information was related to an error or not (see Figure \ref{fig:tensorflowp11}): \textit{``Oh is [sic] like you cannot really tell if there's an error (...) yeah I don't even know if it's the correct output or not'' (P11)}. In another session, while the participant was following installation instructions, the terminal showed the License Terms information. The participant felt bothered by the amount of information and said $-$ \textit{``A lot of text to read...'' (P1)}. Information overload was faced by two out of four participants (\priority{50}) when installing TensorFlow, and also perceived by 51 respondents in our validation survey.

\begin{figure}[h!]
\centering
\includegraphics[width=\linewidth]{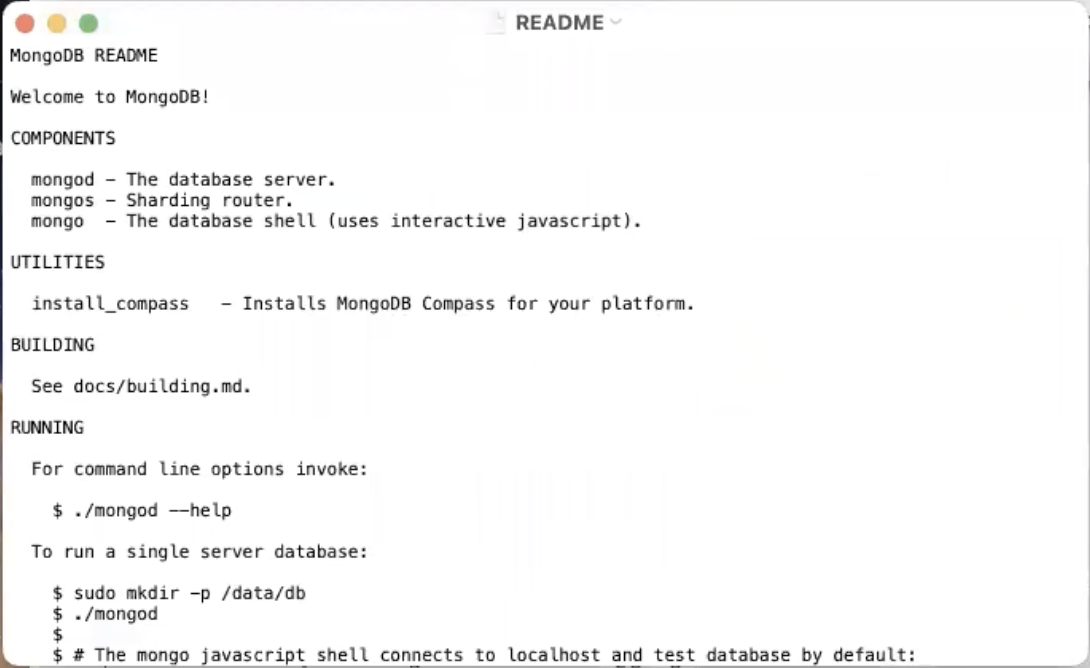}
\caption{Instructions on the README file when installing MongoDB.}
\label{fig:readmep2}
\end{figure}

\underline{\textbf{\textit{Poorly-presented Information}}} refers to the challenges when participants struggled with installation instructions due to the presentation of the documentation.
This challenge occurred in seven sessions when participants were installing Jupyter Notebook, Node.js, Spring Boot, or MongoDB. When installing MongoDB, one of the participants downloaded the package containing executable files. Within the folder, a README file was included (see Figure \ref{fig:readmep2}), which was supposed to provide instructions on how to proceed with the installation. However, the structure of the document made the participant express confusion, stating: \textit{``So... this is quite odd... the README doesn't give much instruction about it'' (P2)}. In another session, while installing Spring Boot, the official website directed a participant to an external site to install a dependency. Upon accessing this site, the participant encountered two buttons labelled ``Download'' and ``Install'', which made them confused. The participant said: \textit{``This is the sort of thing that makes me a little bit confused. It says ``Download'', ``Install'', ``Configure'', ``Run Maven''... so what's the difference between download and install? Yeah, I'm not very sure... '' (P16)}. This challenge was experienced by all participants (\priority[teal!60]{100}) during the installation of Spring Boot and by two out of four participants (\priority{50}) when installing MongoDB. Additionally, 67 survey respondents reported encountering this challenge in the past.

\subsubsection{Installation Process} We identified three challenges that are related to the installation process, i.e., Complicated Installation Process, Lack of Installation Progress Feedback, and Error Message. 

\underline{\textbf{\textit{Complicated Installation Process}}} refers to the challenges when the participants found that there was no ``straightforward'' way to install the tools or that it required too much time.

This challenge commonly occurred in 14 sessions during the installation of all seven tools.
The participant who installed MongoDB was annoyed when there was no direct link to download the tool: \textit{``Now something that I always dislike is there is no direct link to the download part'' (P2)}. The other two participants who installed MongoDB had issues with the download of additional tools. Specifically, they had to install a package manager called Homebrew. Participants found that installing Homebrew and setting up the environment was more complicated and took longer than downloading the tool itself. It took more than four minutes to process each of the commands to install Homebrew: \textit{``I do not know why it is taking so long to update it, should not take that long to update things!'' (P14)}. 

During the installation of PyTorch, the participant wanted to install the tool without downloading any additional tools. The PyTorch website recommends installing the tool using a Python package manager called Anaconda, but the participant wished to find a different installation approach. Thus, the participant searched for ``install PyTorch mac m1 without anaconda'', and when they checked the search results, they realised that they all required Anaconda $-$ \textit{``Everything had Conda on it… Do I need to download Anaconda? Does it come natively?'' (P7)}. The Complicated Installation Process challenge was faced by three out of four participants (\priority{75}) when installing Flutter and MongoDB, and by all participants (\priority[teal!60]{100}) installing Spring Boot. Also, this challenge was selected by 52 respondents in our validation survey.

\underline{\textbf{\textit{Lack of Installation Progress Feedback}}} refers to the challenges when participants did not receive progress feedback when installing the tool, which made participants confused if the installation was successful. This challenge occurred in nine sessions when installing TensorFlow, PyTorch, Jupyter Notebook, Flutter, Node.js, or MongoDB. In the Flutter installation session, when the participant ran one of the installation commands on the terminal, the participant did not receive any feedback or response after waiting for two minutes. A similar problem occurred when installing Jupyter Notebook. This participant followed the instructions from the official website, and when running one of the commands, the terminal stopped showing data processing and nothing happened for three minutes. Participants also expressed confusion when there was no confirmation that the installation process was done. For instance, when the participant executed all the installation instructions of Node.js, they did not receive a confirmation message that the tool was successfully installed: \textit{``I don't know if it's finished or not, how can I check?'' (P12)}. Lack of Installation Progress Feedback was faced by two out of three participants (\priority[teal!60]{67}) during the installation of Jupyter Notebook, Node.js, and two out of four participants (\priority{50}) when installing MongoDB. This challenge was also mentioned by 29 respondents in our validation survey.

\ul{\textbf{\textit{Error Message}}} was encountered when participants faced error messages while attempting to install certain tools. This challenge appeared specifically during the installation process of TensorFlow, Jupyter Notebook, PyTorch, Spring Boot, or MongoDB. For example, during the installation of PyTorch, one participant faced an error message after following instructions from a website. This participant was stuck, grappling with the same error for a long period, lasting more than ten minutes. This challenge was faced by two out of four participants (\priority{50}) installing TensorFlow and MongoDB, respectively, and by two out of three participants (\priority[teal!60]{67}) installing Spring Boot.

\subsubsection{Tool Characteristics} This category represents the challenges that participants faced when they had to deal with the characteristics of the tool they were installing. We identified two challenges that fall under the Tool Characteristics category, i.e., Account Required and Version Incompatibility.

\underline{\textbf{\textit{Account Required}}} refers to the challenges when the tool requires participants to create an account before installing it. This challenge was faced by two out of four participants (\priority{50}) when installing MongoDB. When installing MongoDB, the two participants were looking for installation instructions on MongoDB's official website. When looking for instructions, MongoDB required the user to create an account before having access to the installation instructions. These two participants, when searching for installation instructions, only searched for ``MongoDB'' on Google. In contrast, the other two participants who did not encounter this challenge included ``MongoDB install'' or ``MongoDB installation'' in their queries, which directed them to a different page on MongoDB's official website that did not require account creation. According to our validation survey results, 41 respondents have faced issues related to the Account Required challenge in the past.

\underline{\textbf{\textit{Version Incompatibility}}} refers to challenges when the configuration of the participant's machine is incompatible with the tool they were installing. This challenge occurred in four sessions where participants were installing TensorFlow, PyTorch, Spring Boot, or MongoDB. In all cases, they had incompatibility issues because their machine had the Macbook M1 chip. Version Incompatibility was faced by one participant during the installation of TensorFlow, MongoDB (\priority{25}), PyTorch, and Spring Boot (\priority[teal!60]{33}). Besides, this challenge was mentioned by 30 respondents in our validation survey.

As a result of the ``Other(s)'' category from the survey, respondents pointed out two additional challenges: \textit{``The environment config [sic] is difficult after installation''} and \textit{``The tool required a subscription to be used and the subscription price is expensive''}.






\subsection{Strategies (RQ2)}\label{sec:strategy_results}
Table \ref{tab:strategies} lists 13 strategies that participants applied to overcome the challenges. In this subsection, we organise the strategies into the corresponding challenge categories. See Table \ref{tab:detailedanalisys} for a more detailed analysis of the relationship between challenges, strategies, sources, and tools.

\begin{table}[]
\centering
\caption{Strategies participants applied to overcome the challenges.}
\label{tab:strategies}
\begin{adjustbox}{angle=90,center}
\begin{tabular}{lllcc}
\hline
\textbf{Strategy} & \textbf{Description} & \multicolumn{1}{l}{\textbf{Tools}} & \textbf{No. of Installation Sessions} & \textbf{Survey} \\ \hline
\rowcolor[HTML]{E4E4E4} 
\textbf{Category: Information Quality/Value} & Strategies related to the installation instructions provided to the user &  &  &  \\
\hspace{1em}Searching for Different Documentation & Participant searches for information on a different source & \priority{25} \priority[teal!60]{67} \priority[teal!60]{67} \priority{50} \priority[teal!60]{33} \priority[teal!60]{100} \priority{100} & 15 & 95 \\
\hspace{1em}Reading Help or Feedback in Terminal & Participant uses the terminal instructions to find information & \priority{25} \priority{0} \priority[teal!60]{33} \priority{25} \priority{0} \priority{0} \priority{50} & 5 & 82 \\
\hspace{1em}Reading Instructions Carefully & Participant overcomes the challenge by reading the instructions carefully & \priority{50} \priority{0} \priority[teal!60]{33} \priority{75} \priority[teal!60]{67} \priority[teal!60]{67} \priority{100} & 14 & 83 \\
\rowcolor[HTML]{E4E4E4} 
\textbf{Category: Information Seeking} & Represent information other than official installation instructions &  &  &  \\
\hspace{1em}Reformulating the Query & Participant rephrases the query in order to get more specific results & \priority{50} \priority[teal!60]{33} \priority[teal!60]{33} \priority{25} \priority{0} \priority[teal!60]{33} \priority{100} & 10 & 40 \\
\hspace{1em}Web Searching & Participant searches information (e.g., error messages) on the web & \priority{50} \priority[teal!60]{100} \priority{0} \priority{25} \priority[teal!60]{33} \priority[teal!60]{33} \priority{25} & 9 & 101 \\
\hspace{1em}Focusing on One Source & Participant decides to follow instructions from a single source & \priority{50} \priority[teal!60]{33} \priority[teal!60]{33} \priority{0} \priority[teal!60]{33} \priority[teal!60]{67} \priority{50} & 9 & 28 \\
\rowcolor[HTML]{E4E4E4} 
\textbf{Category: Installation Process} & Events that occurred when participants were in the process of or about to initiate installation &  &  &  \\
\hspace{1em}Giving Up & Participant gives up on installing the tool & \priority{0} \priority[teal!60]{33} \priority{0} \priority{25} \priority{0} \priority[teal!60]{33} \priority{25}& 4 & 16 \\
\hspace{1em}Resetting the Environment & Participant opens a new tab on the terminal & \priority{25} \priority[teal!60]{33} \priority{0} \priority{0} \priority{0} \priority{0} \priority{25} & 3 & 29 \\
\hspace{1em}Reinstalling & Participant decides to restart the installation process & \priority{25} \priority{0} \priority{0} \priority{0} \priority{0} \priority[teal!60]{33} \priority{25} & 3 & 77 \\
\hspace{1em}Waiting & Participant decides to wait for something to happen & \priority{0} \priority{0} \priority[teal!60]{67} \priority{0} \priority[teal!60]{33} \priority{0} \priority{25}& 4 & - \\
\hspace{1em}Random Actions & Participant types random keys/commands & \priority{25} \priority{0} \priority{0} \priority{0} \priority[teal!60]{33} \priority[teal!60]{33} \priority{25} & 4 & - \\
\hspace{1em}Ignoring & Participant ignores error messages or information during the installation process & \priority{0} \priority{0} \priority[teal!60]{33} \priority{25} \priority[teal!60]{67} \priority[teal!60]{67} \priority{50} & 8 & - \\
\rowcolor[HTML]{E4E4E4} 
\textbf{Category: Tool Characteristics} & Comprises the challenges or strategies that occurred due to a particular attribute of the tool &  &  &  \\
\hspace{1em}Versioning & Participant changes or installs a new version of a tool & \priority{25} \priority[teal!60]{33} \priority{0} \priority{0} \priority{0} \priority{0} \priority{0} & 2 & 48 \\ \hline
\multicolumn{5}{l}{\textbf{Note:} The order of the tools is: TensorFlow, PyTorch, Jupyter Notebook, Flutter, Node.js, Spring Boot, and MongoDB.} \\ \hline
\end{tabular}
\end{adjustbox}
\end{table}


\subsubsection{Information Quality/Value}
In this category, we identified three strategies, i.e., Searching for Different Documentation, Reading Help or Feedback in Terminal, and Reading Instructions Carefully.

\underline{\textbf{\textit{Searching for Different Documentation}}} was used in 15 sessions when installing all seven tools. Stack Overflow was one of the preferred alternative sources that the participants used: \textit{``For most things, yes. If it's a Python package I just look for the PIP or the keyword for PIP, but other than that I would go to Stack Overflow'' (P6)}. Similarly, in the session of installing PyTorch, the participants attempted to follow the instructions on the tool's official website. However, the participants perceived that the instructions there were not helpful. Then, they decided to follow the installation instructions in the non-official documentation, as the participants perceived it was easier to understand than the official documentation: \textit{``Yes. I prefer this because it has more explanation and it is more user-friendly'' (P10)}. Out of 15 sessions, looking for alternative sources helped the participants overcome the challenges in three sessions. In eight out of 15 sessions, the participants preferred non-official documentation over the tool's official documentation. Similarly, participants often preferred to watch video tutorials over reading extensive documents: \textit{``I usually just look for them in the software documentation, but sometimes it's clearer to just see [sic] a video, like maybe a five-minute walkthrough. If I get stuck I usually go watch a video'' (P8)}.

While some participants preferred non-official documentation, other participants found that such documentation can be outdated.
In one of the sessions of installing MongoDB, the participant first followed the installation instructions from a non-official website. When faced with challenges, they searched for information in the official documentation as they believed that it was a more reliable source:\textit{``The official website is more recent, and if there is some sort of update, some other discussion forum would be outdated.'' (P14)}. Nevertheless, when participants faced an installation error, reading the official documentation was not helpful to the participants as it did not provide a solution for such an error. This strategy was applied to the challenges: \ul{\textit{Unclear Instructions}}, \ul{\textit{Dysfunctional Instructions, Under-specified query, Error Message, Complicated Installation Process, Lack of Installation Progress Feedback, Account Required}}, and \ul{\textit{Version Incompatibility}}. Searching for Different Documentation was applied by two out of three participants (\priority[teal!60]{67}) when installing PyTorch or Jupyter Notebook, as well as by all participants when installing Spring Boot (\priority[teal!60]{100}) and MongoDB (\priority{100}). Additionally, 95 respondents in our validation survey mentioned having applied this strategy in the past when installing software development tools.

\underline{\textbf{\textit{Reading Help or Feedback in Terminal}}} was used when participants tried to understand the installation instructions. This strategy was used in five sessions when installing TensorFlow, Jupyter Notebook, Flutter, or MongoDB. During the session of installing MongoDB, participants struggled to understand how to run the installation command as per the instructions from the official documentation. Then, one of the participants was looking for help on the shell's documentation using the ``--help'' command. Similarly, in the TensorFlow installation session, the participant had issues with an error message when executing a command. One of their strategies was to carefully read all the information shown on the terminal. Reading Help or Feedback in Terminal was applied to the challenges: \underline{\textit{Unclear Instructions}}, \underline{\textit{Dysfunctional Instructions}} and \ul{\textit{Information Overload}}. This strategy was applied by two out of four participants (\priority{50}) during the installation of MongoDB, as well as by one participant when installing TensorFlow, Flutter (\priority{25}), or Jupyter Notebook (\priority[teal!60]{33}). In our validation survey, 82 respondents highlighted the use of this strategy to overcome challenges.


\underline{\textbf{\textit{Reading Instructions Carefully}}} was used when the participants could not understand or follow the installation instructions. Initially, participants tended to skim the information, but once they encountered the challenge, they returned to the same information but read it word by word. For example, when installing Spring Boot, one participant first glanced through the installation instructions and claimed there was no direct download link. However, after slowing down and reading the page more carefully, they realised that there was a download link on the page. This strategy was used in 14 sessions where participants were installing TensorFlow, Jupyter Notebook, Flutter, Node.js, Spring Boot, or MongoDB. This strategy successfully helped the participants overcome the challenge once. This strategy was applied to the challenges: \ul{\textit{Unclear Instructions, Dysfunctional Instructions, Poorly-structured Information, Information Overload, Complicated Installation Process, Lack of Installation Progress Feedback}}, and \ul{\textit{Version Incompatibility}}. Reading Instructions Carefully was applied by two out of three participants (\priority[teal!60]{67}) who installed Node.js or Spring Boot, as well as by two out of three participants (\priority[teal!60]{67}) during the Spring Boot installation, and by all participants (\priority{100}) who installed MongoDB. In total, 83 respondents reported applying this strategy previously.

\subsubsection{Information Seeking}
In this category, we identified three strategies (i.e., Reformulating the Query, Web Searching, and Focusing on One Source).

\underline{\textbf{\textit{Reformulating the Query}}} was used in 10 sessions when installing TensorFlow, PyTorch, Jupyter Notebook, Flutter, Spring Boot, or MongoDB. As pointed out in the challenge of Under-specified Query, the initial search keyword was not specific (e.g., using only the tool name as a keyword search). Thus, the participants attempted to overcome this challenge by reformulating the query and adding more specifications or context, such as the operating system they were using. For instance, when installing MongoDB, one of the participants used only ``MongoDB'' as a keyword to search on Google and did not find any useful links. The participant then realised that additional keywords should be used to retrieve more useful results: \textit{``I couldn't find the download directly from that page, so I had to go back and search for ``download MongoDB''. Then, I downloaded something that did not work, and then, by making a full query of like ``installing MongoDB on MacOS'', then I got the instructions to use, like the command line with Homebrew, and then it worked'' (P2)}. This strategy was applied to the challenges: \ul{\textit{Unclear Instructions, Dysfunctional Instructions, Under-specified Query, Poorly-structured Information, Error Message}}, and \ul{\textit{Complicated Installation Process}}. Reformulating the Query was applied by two out of four participants (\priority{50}) when installing TensorFlow, and by all participants (\priority{100}) who installed MongoDB. In our validation survey, 40 respondents mentioned using this strategy as a means of overcoming challenges.


\textbf{\underline{\textit{Web Searching}}} was employed when participants needed to find information online to address the challenges they encountered. This strategy was applied in nine sessions across the installation of TensorFlow, PyTorch, Flutter, Node.js, Spring Boot, or MongoDB. In two sessions involving TensorFlow, participants encountered error messages during installation and chose to search for the error messages online as a strategy. In three other sessions, when installing TensorFlow, Node.js, or MongoDB, participants experienced a lack of installation progress feedback. To address this challenge, they searched the web to determine whether the issue was common and to find possible solutions. This strategy was applied to the challenges: \textit{\ul{Complicated installation process, Lack of Installation Progress Feedback, Error Message, Dysfunctional Instructions, Unclear instructions, and Information Overload.}} Web Searching was applied by two out of four participants (\priority{50}) installing TensorFlow and by all three participants (\priority[teal!60]{100}) installing PyTorch. Besides, 101 participants in our validation survey reported having applied this challenge in the past.

\underline{\textbf{\textit{Focusing on One Source}}} was used when the participant used installation instructions on multiple sources at the same time, which turned out to be problematic due to inconsistent instructions. To overcome that challenge, the participant decided to follow the instructions of one source only, and that solved the problem. This strategy was applied in nine sessions during the TensorFlow, PyTorch, Jupyter Notebook, Node.js, Spring Boot, or MongoDB installation process. This strategy was applied to the challenges: \ul{\textit{Dysfunctional Instructions, Under-specified Query, Poorly-structured Information, Error Message, Complicated Installation Process, Lack of Installation Progress Feedback}}, and \ul{\textit{Account Required}}. Focusing on One Source was applied by two out of three participants (\priority[teal!60]{67})  installing Spring Boot and by two out of four participants (\priority{50}) installing MongoDB or TensorFlow. Also, this strategy was reported by 28 respondents in our validation survey.

\subsubsection{Installation Process} 
In this category, we identified six strategies (i.e., Giving Up, Resetting the Environment, Reinstalling, Waiting, Random Actions, and Ignoring).

\underline{\textbf{\textit{Giving Up}}} was applied when participants felt exhausted after spending a considerable amount of time attempting to install the tool. One of the participants mentioned: \textit{``I'm gonna give up, the unzipping and extracting files process is taking too long'' (P6)}. In addition, some participants gave up on installing the tool because they wanted to install it without having to install additional tools. This strategy occurred in four sessions when installing PyTorch, Flutter, Spring Boot, or MongoDB. This strategy was applied to the challenges: \ul{\textit{Dysfunctional Instructions, Poorly-structured Information, Error Message, Complicated Installation Process}}, and \ul{\textit{Version Incompatibility}}. In total, one participant decided to give up during the installation process of PyTorch, Spring Boot (\priority[teal!60]{33}), Flutter, or MongoDB (\priority{25}). Additionally, 16 respondents in our validation survey mentioned using this strategy to address challenges they encountered previously.


\underline{\textbf{\textit{Resetting the Environment}}} was applied when participants faced errors and could not understand the problem due to a lack of feedback on installation progress when installing TensorFlow, PyTorch, or MongoDB. In the three sessions, the participants attempted to resolve the problem by resetting the environment (i.e., opening a new terminal tab) and found that this strategy was helpful. This strategy was applied to the challenge: \ul{\textit{Lack of Installation Progress Feedback}}. During the installation sessions, this strategy was applied by three out of four participants (\priority{75}) when installing TensorFlow or MongoDB, and by all three participants (\priority[teal!60]{100}) who installed PyTorch. According to our validation survey results, this strategy was selected by 29 respondents.

\underline{\textbf{\textit{Reinstalling}}} was another strategy used to overcome challenges in three sessions when installing TensorFlow, Spring Boot or MongoDB. This strategy was mostly unsuccessful due to either dysfunctional instructions or a lack of installation progress feedback. When installing MongoDB, one participant encountered an issue where no installation progress was displayed and attempted to reinstall the tool twice, but this did not resolve the problem. During the installation of TensorFlow, the instructions required the participant to install a dependency, but their computer failed to recognise the installation. To address this, they decided to uninstall and reinstall the tool, which ultimately resolved the issue. In another session, a participant installing Spring Boot encountered an unexpected error message. After multiple unsuccessful attempts to resolve the issue, they decided that reinstalling the tool could be a viable strategy. Although this approach was applied in three sessions, it was only successful in the TensorFlow installation. This strategy was applied to the challenges: \ul{\textit{Dysfunctional Instructions, Lack of Installation Progress Feedback, and Error Message}}. Moreover, this strategy was applied by one out of four participants (\priority{25}) during the installation of TensorFlow or MongoDB, and by one out of three (\priority[teal!60]{33}) participants when installing Spring Boot. In our validation survey, 77 respondents revealed having applied this strategy previously.

\ul{\textbf{\textit{Waiting}}} was an option for the participants in four sessions when installing Jupyter Notebook, Node.js or MongoDB. In one of the sessions, the participant installing MongoDB experienced a waiting period of more than 10 minutes before receiving any feedback on the progress of the installation. When installing Node.js, one participant did not receive any updates for over four minutes, however, they decided that waiting was the best strategy. This approach proved successful for all four participants across the sessions. This strategy was applied to the challenge \ul{\textit{Lack of Installation Progress Feedback}}. The strategy ``Waiting'' was applied by two out of the three participants (\priority[teal!60]{67}) who installed Jupyter Notebook.

\ul{\textbf{\textit{Random Actions}}} were taken by three participants during the installation process of TensorFlow, Node.js, Spring Boot, or MongoDB. When installing TensorFlow, the participant decided to click on random keys on the keyboard, this choice was motivated by the absence of feedback regarding the installation progress. This strategy worked in this case, but was not successful in other sessions. For example, when installing Node.js, one participant did not receive any feedback on the progress of the Python installation. The only message displayed was ``Installing 64-bit python311...'', but the screen remained stuck on this message for three minutes. This prompted them to press the ``Enter'' key multiple times in an attempt to get a response from the system, but nothing happened. This strategy was applied to the challenges: \ul{\textit{Lack of Installation Progress Feedback}} and \ul{\textit{Unclear Instructions}}. Additionally, one participant out of four (\priority{25}) applied this strategy when installing TensorFlow or MongoDB, and one out of three participants (\priority[teal!60]{33}) during the installation of Node.js or Spring Boot.


\ul{\textbf{\textit{Ignoring}}} was employed in eight sessions when installing Jupyter Notebook, Flutter, Node.js, Spring Boot, or MongoDB. In one of the sessions, the participant was following the installation instructions from the official Spring Boot website. The official website was perceived as less user-friendly, making the installation instructions somewhat difficult to follow. Consequently, this participant opted to ignore certain installation steps described on the website. This strategy was applied to the challenges: \ul{\textit{Unclear Instructions, Poorly-structured Information, Account Required}}, and \ul{\textit{Error Message}}. Ignoring was applied by two out of three participants (\priority[teal!60]{67}) during the installation of Node.js, Spring Boot, and by two out of four participants (\priority{50}) when installing MongoDB.



\subsubsection{Tool Characteristics}

\underline{\textbf{\textit{Versioning}}} is the only strategy of the Tool Characteristics category. It was applied by two participants, one in each session when installing TensorFlow (\priority{25}) or PyTorch (\priority[teal!60]{33}). In both cases, the participants were facing issues installing the tools and were trying to change the Python version to see if the issue could be solved. This strategy was successful during the installation of TensorFlow, but it was not successful when the participant installed Python. This strategy was applied to the challenges: \ul{\textit{Error Message}} and \ul{\textit{Version Incompatibility}}. This strategy was selected by 48 respondents in our validation survey.

Regarding the ``Other(s)'' option in our validation survey, we only had one single addition for the strategies - \textit{``call my excellent [sic] friend for help''}.

\subsection{Sources (RQ3)}

Previously, we explained each challenge and strategy by providing examples extracted from the think-aloud sessions in situations where these challenges and strategies were perceived and applied. In this subsection, we present the sources that were consulted when applying each strategy and the tools in which those sources were applied during the 24 think-aloud installation sessions. Moreover, we present a detailed analysis of the relationship between challenges and strategies, introducing the type of sources (see Table \ref{tab:sources}) consulted by the participants when applying the strategies.  For each challenge (see Table \ref{tab:detailedanalisys}), we document the specific strategies applied, the sources consulted during that period, and the tools for which those sources were accessed. In this table, we use the challenge categories to guide its interpretation. So, the categories presented are strictly related to the challenges and not to the strategies.


 We mapped and categorised the sources consulted during the installation sessions. The sources were mapped based on the specific information the participant was consulting at a given moment. For example, if a participant was reading an installation guide that included a video tutorial, we only considered the content they were actively engaging with. To facilitate the categorisation process, we initially adopted a terminology consistent with previous research \citep{aghajani2020software}, such as Getting Started, Installation Guide, Video tutorials, How-To/Tutorial, and Community Knowledge. Regarding the sources that did not align with any pre-existing categories established in the prior work, we formulated new categories: Dashboard, Sign Up Page, Google Results, Download Page, Terminal, README, and No Source. In Table \ref{tab:sources}, each source category is represented by a specific symbol. The Community Knowledge (\textcolor{MidnightBlue}{\faUsers}) source encompasses online community forums (e.g., Stack Overflow), How-To/Tutorial (\textcolor{myblue}{\faPencilSquare}) refers to general guides, typically in the form of blogs. Video tutorials (\textcolor{red}{\faYoutubePlay}) comprise instructional videos illustrating tool installation procedures.

 \begin{table}[H]
\centering
\caption{Sources type we used for the detailed analysis.}
\tiny
\label{tab:sources}\resizebox{\textwidth}{!}{
\begin{tabular}{@{}lcllclcllcl@{}}
\toprule
\textbf{} & \multicolumn{1}{l}{\textbf{Source Symbol}} &  & \textbf{Source Name} & \textbf{Occ.} & \textbf{} & \multicolumn{1}{l}{\textbf{Source Symbol}} &  & \textbf{Source Name} & \textbf{Occ.} & \textbf{} \\ \midrule
 & \textcolor{MidnightBlue}{\faUsers} &  & Community Knowledge & 18 &  & \textcolor{CadetBlue}{\faSignIn} &  & Sign Up Page & 2 &  \\
 & \textcolor{myblue}{\faPencilSquare}  &  & How-To Tutorial & 20 &  & \textcolor{RoyalBlue}{\faGoogle} &  & Google Results & 3 &  \\
 & \textcolor{red}{\faYoutubePlay} &  & Video Tutorials & 7 &  & \color{PineGreen} \faDownload &  & Download Page & 6 &  \\
 & \textcolor{gray}{\faGear} &  & Installation Guide & 55 &  & \textcolor{darkgray}{\faIndent} &  & Terminal & 19 &  \\
 & \textcolor{myyellow}{\faThumbsUp} &  & Getting Started & 38 &  & \color{mygray} \faFileText &  & README & 1 &  \\
 & \textcolor{mybrown}{\faHome} &  & Dashboard & 6 &  & \textcolor{purple}{\faTimesCircle} &  & No Source & 13 &  \\ 
  & \textcolor{SeaGreen}{\faCommenting} &  & ChatGPT & 2 &  &  &  &  &  &  \\
\bottomrule
\end{tabular}}
\end{table}

\vspace{-30pt}
\begin{table}[H]
\centering
\caption{Detailed analysis of each challenge, the applied strategy and the consulted source during the installation of the specific tools.}
\label{tab:detailedanalisys}
\resizebox{\textwidth}{!}{
\begin{tabular}{llll}
\hline
\rowcolor[HTML]{DADADA} \multicolumn{4}{c}{\cellcolor[HTML]{E4E4E4}\textbf{Challenge Category: Information Quality/Value}} \\ \hline
\multicolumn{2}{c|}{\textbf{Challenge: Unclear Instructions}} & \multicolumn{2}{c}{\textbf{Challenge: Dysfunctional Instructions}} \\ \hline
\textbf{Strategies} & \textbf{Sources} & \textbf{Strategies} & \textbf{Sources} \\ \hline
Searching for Different Documentation & \textcolor{mybrown}{\faHome} $_\textit{M}$ \textcolor{myblue}{\faPencilSquare} $_\textit{F}$  \textcolor{gray}{\faGear} $_\textit{F, N}$ \textcolor{myyellow}{\faThumbsUp} $_\textit{N, S}$ \textcolor{PineGreen}{\faDownload} $_\textit{S}$ & Searching for Different Documentation & \textcolor{MidnightBlue}{\faUsers} $_\textit{M}$ \textcolor{red}{\faYoutubePlay} $_\textit{S}$ \textcolor{gray}{\faGear} $_\textit{J, N}$ \textcolor{myyellow}{\faThumbsUp} $_\textit{S, N}$ \\ Reading Help or Feedback in Terminal & \textcolor{darkgray}{\faIndent} $_\textit{M, F}$ & Reading Help or Feedback in Terminal & \textcolor{gray}{\faGear} $_\textit{J}$ \textcolor{darkgray}{\faIndent} $_\textit{F}$ \\
Reading Instructions Carefully & \textcolor{mygray} {\faFileText}  $_\textit{M}$ \textcolor{darkgray}{\faIndent} $_\textit{N}$ \textcolor{gray}{\faGear} $_\textit{J, S}$ & Reading Instructions Carefully & \textcolor{MidnightBlue}{\faUsers} $_\textit{M}$ \textcolor{gray}{\faGear} $_\textit{F, J, S}$ \textcolor{myyellow}{\faThumbsUp} $_\textit{F, S}$ \\
Focusing on One Source & \textcolor{gray}{\faGear} $_\textit{J}$ \textcolor{myyellow}{\faThumbsUp} $_\textit{S}$ \textcolor{purple}{\faTimesCircle} $_\textit{S}$ & Focusing on One Source & \textcolor{MidnightBlue}{\faUsers} $_\textit{M}$ \textcolor{gray}{\faGear} $_\textit{S}$ \textcolor{myyellow}{\faThumbsUp} $_\textit{S}$ \\ 
Reformulating the Query & \textcolor{MidnightBlue}{\faUsers} $_\textit{T}$  \textcolor{PineGreen}{\faDownload} $_\textit{S}$ & Reformulating the Query & \textcolor{myyellow}{\faThumbsUp} $_\textit{S}$ \textcolor{red}{\faYoutubePlay} $_\textit{M}$ \\
Web Searching & \textcolor{myblue}{\faPencilSquare} $_\textit{M}$ \textcolor{gray}{\faGear} $_\textit{M, N}$ & Web Searching & \textcolor{MidnightBlue}{\faUsers} $_\textit{T}$ \textcolor{red}{\faYoutubePlay} $_\textit{M}$ \\ 
Random Actions & \textcolor{darkgray}{\faIndent} $_\textit{M}$ \textcolor{myyellow}{\faThumbsUp} $_\textit{S}$ & 
Giving Up & \textcolor{purple}{\faTimesCircle} $_\textit{S}$ \\ Reinstalling & \textcolor{MidnightBlue}{\faUsers} $_\textit{T, M}$ \textcolor{gray}{\faGear} $_\textit{S}$ & Ignoring & \textcolor{darkgray}{\faIndent} $_\textit{J}$ \\
Ignoring & \textcolor{gray}{\faGear} $_\textit{S}$ \textcolor{myyellow}{\faThumbsUp} $_\textit{S}$ \textcolor{darkgray}{\faIndent} $_\textit{N}$ \\\hline
\rowcolor[HTML]{DADADA} \multicolumn{4}{c}{\cellcolor[HTML]{E4E4E4}\textbf{Challenge Category: Information Seeking}} \\ \hline
\multicolumn{2}{c|}{\textbf{Challenge: Under-specified Query}} & \multicolumn{2}{c}{\textbf{Challenge: Poorly-structured Information}} \\ \hline
\textbf{Strategies} & \textbf{Sources} & \textbf{Strategies} & \textbf{Sources} \\ \hline
Searching for Different Documentation & \textcolor{myblue}{\faPencilSquare} \textcolor{gray}{\faGear} \textcolor{myyellow}{\faThumbsUp} $_\textit{P}$ & Reading Instructions Carefully & \textcolor{gray}{\faGear} $_\textit{J}$ \textcolor{darkgray}{\faIndent} \textcolor{purple}{\faTimesCircle} $_\textit{N}$ \\
Focusing on One Source & \textcolor{RoyalBlue} {\faGoogle} $_\textit{N}$ & Focusing on One Source & \textcolor{gray}{\faGear} $_\textit{J, M}$ \textcolor{myyellow}{\faThumbsUp} $_\textit{M}$  \textcolor{red}{\faYoutubePlay} $_\textit{S}$ \\
Reformulating the Query & \begin{tabular}[t]{@{}l@{}}\textcolor{myblue}{\faPencilSquare} $_\textit{P}$ \textcolor{gray}{\faGear} \textit{$_\textit{M, P}$} \textcolor{myyellow}{\faThumbsUp} \textit{$_\textit{T, P}$} \textcolor{CadetBlue}{\faSignIn} $_\textit{M}$ \textcolor{RoyalBlue} {\faGoogle} $_\textit{J}$ \end{tabular} & Reformulating the Query & \textcolor{myblue}{\faPencilSquare} \textcolor{gray}{\faGear} $_\textit{M}$ \\
Web Searching & \textcolor{myblue}{\faPencilSquare} \textcolor{gray}{\faGear} \textcolor{myyellow}{\faThumbsUp} $_\textit{P}$ & Random Actions & \textcolor{myyellow}{\faThumbsUp} $_\textit{S}$ \\ \textbf{} & \textbf{} & Ignoring & \textcolor{gray}{\faGear} $_\textit{S}$ \\\hline
\multicolumn{2}{c|}{\textbf{Challenge: Information Overload}} & \multicolumn{2}{c}{\textbf{Challenge: Error Message}} \\ \hline
\textbf{Strategies} & \textbf{Sources} & \textbf{Strategies} & \textbf{Sources} \\ \hline
Reading Instructions Carefully & \textcolor{darkgray}{\faIndent} $_\textit{T}$ & Searching for Different Documentation & \textcolor{myblue}{\faPencilSquare} \textcolor{gray}{\faGear} \textcolor{myyellow}{\faThumbsUp} $_\textit{P}$ \textcolor{gray}{\faGear} $_\textit{J}$ \textcolor{PineGreen}{\faDownload} $_\textit{S}$ \\ Reading Help or Feedback in Terminal & \textcolor{darkgray}{\faIndent} $_\textit{T}$ & Reading Help or Feedback in Terminal & \textcolor{gray}{\faGear} $_\textit{T}$ \\ Web Searching & \textcolor{RoyalBlue} {\faGoogle} $_\textit{T}$ & Reading Instructions Carefully & \textcolor{gray}{\faGear} $_\textit{S, J}$ \\ Ignoring & \textcolor{darkgray}{\faIndent} $_\textit{N, M}$ & Web Searching & \textcolor{MidnightBlue}{\faUsers} $_\textit{P, N, T}$ \\
 & & Focusing on One Source & \textcolor{gray}{\faGear} \textcolor{myyellow}{\faThumbsUp} $_\textit{P}$ \textcolor{red}{\faYoutubePlay} $_\textit{M}$ \\
& & Reformulating the Query & \textcolor{darkgray}{\faIndent} $_\textit{M}$ \\
 &  & Giving Up & \textcolor{gray}{\faGear} \textcolor{myyellow}{\faThumbsUp} $_\textit{P}$ \\
 &  & Versioning & \textcolor{darkgray}{\faIndent} $_\textit{P}$ \\ & & Reinstalling & \textcolor{gray}{\faGear} $_\textit{S}$ \\ & & Ignoring & \textcolor{gray}{\faGear} \textcolor{darkgray}{\faIndent} \textcolor{SeaGreen}{\faCommenting} $_\textit{J}$ \textcolor{myblue}{\faPencilSquare} \textcolor{PineGreen}{\faDownload} $_\textit{M}$ \\ \hline
\rowcolor[HTML]{DADADA} \multicolumn{4}{c}{\cellcolor[HTML]{E4E4E4}\textbf{Challenge Category: Installation Process}} \\ \hline
\multicolumn{2}{c|}{\textbf{Challenge: Complicated Installation Process}} & \multicolumn{2}{c}{\textbf{Challenge: Lack of Installation Progress Feedback}} \\ \hline
\textbf{Strategies} & \textbf{Sources} & \textbf{Strategies} & \textbf{Sources} \\ \hline
Searching for Different Documentation & \textcolor{MidnightBlue}{\faUsers} $_\textit{T, F, S}$ \textcolor{myblue}{\faPencilSquare} $_\textit{P}$ \textcolor{gray}{\faGear} $_\textit{T, P, N}$ \\ &  \textcolor{myyellow}{\faThumbsUp} $_\textit{P, N, S}$ \textcolor{PineGreen}{\faDownload} $_\textit{N}$ \textcolor{red}{\faYoutubePlay} $_\textit{S}$ \textcolor{mybrown}{\faHome} $_\textit{M}$  & Searching for Different Documentation & \textcolor{red}{\faYoutubePlay} $_\textit{F}$ \textcolor{myyellow}{\faThumbsUp} $_\textit{M}$ \\
Reading Instructions Carefully & \textcolor{myblue}{\faPencilSquare} $_\textit{F, S, M}$ \textcolor{gray}{\faGear} $_\textit{F, S}$ \textcolor{CadetBlue}{\faSignIn} $_\textit{M}$ \textcolor{myyellow}{\faThumbsUp} $_\textit{F}$ \textcolor{PineGreen}{\faDownload} $_\textit{M}$ & Reading Instructions Carefully & \textcolor{darkgray}{\faIndent} $_\textit{M}$  \\
Focusing on One Source & \textcolor{MidnightBlue}{\faUsers} $_\textit{T}$ \textcolor{myblue}{\faPencilSquare} $_\textit{T, M}$ \textcolor{gray}{\faGear} $_\textit{T, M}$ \textcolor{mybrown}{\faHome} $_\textit{M}$ \textcolor{myyellow}{\faThumbsUp} $_\textit{S}$ \textcolor{SeaGreen}{\faCommenting} $_\textit{J}$ & Focusing on One Source & \textcolor{red}{\faYoutubePlay} $_\textit{M}$ \textcolor{myyellow}{\faThumbsUp} $_\textit{J}$ \\
Reformulating the Query & \textcolor{MidnightBlue}{\faUsers} $_\textit{F}$ \textcolor{gray}{\faGear} $_\textit{P, J}$ \textcolor{mybrown}{\faHome} $_\textit{M}$ & Web Searching & \textcolor{red}{\faYoutubePlay} $_\textit{M}$ \textcolor{myblue}{\faPencilSquare} \textcolor{gray}{\faGear} $_\textit{P}$  \\
Web Searching & \textcolor{MidnightBlue}{\faUsers} $_\textit{F}$ \textcolor{myblue}{\faPencilSquare} \textcolor{gray}{\faGear} \textcolor{myyellow}{\faThumbsUp} $_\textit{P}$ \textcolor{mybrown}{\faHome} $_\textit{S}$ & Resetting the Environment & \textcolor{purple}{\faTimesCircle} $_\textit{M}$ \\
Giving Up & \textcolor{purple}{\faTimesCircle} $_\textit{F}$ & Random Actions & \textcolor{purple}{\faTimesCircle} $_\textit{M}$ \textcolor{darkgray}{\faIndent} $_\textit{N}$ \\ Ignoring & \textcolor{myyellow}{\faThumbsUp} \textcolor{gray}{\faGear} $_\textit{S, F}$ & Waiting & \textcolor{purple}{\faTimesCircle} $_\textit{M, J}$ \textcolor{darkgray}{\faIndent} $_\textit{N}$ \\
 &  & Reinstalling & \textcolor{purple}{\faTimesCircle} $_\textit{M}$ \\
 &  & Ignoring & \textcolor{darkgray}{\faIndent} $_\textit{M}$ \\ \hline
\rowcolor[HTML]{DADADA} \multicolumn{4}{c}{\cellcolor[HTML]{E4E4E4}\textbf{Challenge Category: Tool Characteristics}} \\ \hline
\multicolumn{2}{c|}{\textbf{Challenge: Account Required}} & \multicolumn{2}{c}{\textbf{Challenge: Version Incompatibility}} \\ \hline
\textbf{Strategies} & \textbf{Sources} & \textbf{Strategies} & \textbf{Sources} \\ \hline
Searching for Different Documentation & \textcolor{mybrown}{\faHome} $_\textit{M}$ & Searching for Different Documentation & \textcolor{MidnightBlue}{\faUsers} $_\textit{M}$ \textcolor{gray}{\faGear} \textcolor{myyellow}{\faThumbsUp} $_\textit{P}$ \\
Focusing on One Source & \textcolor{mybrown}{\faHome} $_\textit{M}$ & Reading Instructions Carefully & \textcolor{gray}{\faGear} \textcolor{myyellow}{\faThumbsUp} $_\textit{T}$ \\
& & Giving Up & \textcolor{MidnightBlue}{\faUsers} $_\textit{M}$ \\
 &  & Reinstalling & \textcolor{gray}{\faGear} \textcolor{myyellow}{\faThumbsUp} $_\textit{T}$  \\
 &  & Versioning & \textcolor{gray}{\faGear} \textcolor{myyellow}{\faThumbsUp} $_\textit{T, P}$ \\ 
&  & Ignoring & \textcolor{gray}{\faGear} \textcolor{myyellow}{\faThumbsUp} $_\textit{S}$ \\ \hline
\multicolumn{4}{l}{\small \textit{T = TensorFlow | F = Flutter | M = MongoDB | P = PyTorch | J = Jupyter Notebook | N = Node.js | S = Spring Boot}} \\ \bottomrule
\end{tabular}}
\end{table}

Installation Guide (\textcolor{gray}{\faGear}) denotes comprehensive, step-by-step documents or web pages providing installation instructions, while Getting Started (\textcolor{myyellow}{\faThumbsUp}) refers to materials offering an introductory overview of the tool and its installation process. The Dashboard (\textcolor{mybrown}{\faHome}) category refers to the homepage of the tool’s official website. ChatGPT (\textcolor{SeaGreen}{\faCommenting}) represents instances where participants consulted ChatGPT as a source. Sign Up Page (\textcolor{CadetBlue}{\faSignIn}) indicates the registration page on the official tool website. The category Google Results (\textcolor{RoyalBlue}{\faGoogle}) represents findings from a Google search, and Download Page (\textcolor{PineGreen}{\faDownload}) designates a webpage facilitating the installation process (e.g., a .exe file). Information originating from the terminal prompt is categorised as Terminal (\textcolor{darkgray}{\faIndent}). README (\textcolor{mygray}{\faFileText}) represents documentation provided in README files, and No Source (\textcolor{purple}{\faTimesCircle}) is used when no external source was consulted.





\textbf{Community Knowledge} \textcolor{MidnightBlue}{\faUsers} emerged as a frequently consulted source, particularly for the challenges Unclear Instructions, Dysfunctional Instructions, Error Message, Complicated Installation Process, and Version Incompatibility. This source was mainly consulted when employing the strategies Searching for Different Documentation and Web Searching. The Community Knowledge source was applied when installing TensorFlow, MongoDB and Flutter. These findings show that online forums play an important role in supporting novice developers to overcome challenges during the software development tool installation process.

\textbf{How-To/Tutorial} \textcolor{myblue}{\faPencilSquare} sources were used in six challenges, mainly addressing the challenges of Unclear Instructions, Under-specified Query, and Poorly-structured Information. Similar to the Community Knowledge source, these were accessed primarily while implementing the strategies of Searching for Different Documentation, Web Searching, and Focusing on One Source. This source was frequently accessed when installing PyTorch, MongoDB, and TensorFlow. 

\textbf{Video Tutorials} \textcolor{red}{\faYoutubePlay} were consulted for five challenges, including Dysfunctional Instructions, Error Message, and Lack of Installation Progress Feedback. This source was used in conjunction with the strategies Searching for Different Documentation, Web Searching, and Focusing on One Source, particularly during the installation process of MongoDB, Flutter, and Spring Boot.

\textbf{Installation Guide} \textcolor{gray}{\faGear} proved to be the most frequently used source, being employed for eight challenges in total, but mostly for Unclear Instructions, Dysfunctional Instructions, Under-specified Query, Poorly-structured Information, Error Message, Complicated Installation Process, and Version Incompatibility. Participants consulted this source when applying six strategies, for the most part, Searching for Different Documentation, Reading Instructions Carefully and Web Searching. This demonstrates the utility of step-by-step installation guides for novice developers who navigate the tool installation process. The Installation Guide source was mostly consulted when installing PyTorch, TensorFlow, Flutter, Node.js, and MongoDB.

\textbf{Getting Started} \textcolor{myyellow}{\faThumbsUp} was applied to eight challenges, including Under-specified Query and Version Incompatibility. It was associated with a total of seven strategies, with Searching for Different Documentation, Focusing on One Source, and Reading Instructions Carefully being the most common. This source was accessed during the installation process of Flutter, Node.js, MongoDB, PyTorch, TensorFlow, and Jupyter Notebook.

\textbf{Dashboard} \textcolor{mybrown}{\faHome} was comparatively less utilised, featuring only three challenges, specifically Unclear Instructions, Complicated Installation Process, and Account Required. This source was accessed primarily during the installation of MongoDB and Spring Boot and was correlated with strategies such as Searching for Different Documentation and Focusing on One Source.

There are four sources with fewer occurrences. The \textbf{ChatGPT} \textcolor{SeaGreen}{\faCommenting} source was consulted only twice, both times by the same participant. It was used once during the Error Message challenge when the strategy Ignoring was applied, and once during the Complicated Installation Process challenge when the strategy Focusing on One Source was applied. \textbf{Sign Up Page} \textcolor{CadetBlue}{\faSignIn} source was applied in Under-specified Query and Complicated Installation Process, with the strategy Reading Instructions Carefully being employed. This source was consulted during the MongoDB installation. \textbf{Google Results} \textcolor{RoyalBlue}{\faGoogle} was accessed for Under-specified Query and Information Overload challenges, with strategies such as Focusing on One Source, Reformulating the Query, and Web Searching. These instances were associated with the installations of Node.js, Jupyter Notebook, and TensorFlow. \textbf{Download Page} \textcolor{PineGreen}{\faDownload} source was consulted in three challenges, Unclear Instructions, Error Message, Complicated Installation Process, and was accessed primarily during the strategy Searching for Different Documentation during the installation process of Node.js.

\textbf{Terminal} \textcolor{darkgray}{\faIndent} source was used in six challenges, mainly addressing issues such as Unclear Instructions, Dysfunctional Instructions, and Lack of Installation Progress Feedback. At the same time, this source was consulted in connection with five strategies, with a focus on Reading Help or Feedback in Terminal and Reading Instructions Carefully. These strategies were predominantly applied during the installation processes of Flutter, Node.js, MongoDB, PyTorch, and TensorFlow, indicating that the participants found valuable information in the terminal output. 

\textbf{README} \textcolor{mygray}{\faFileText} source made a single appearance in the challenge Unclear Instructions, primarily employed in the Reading Instructions Carefully strategy, and was consulted with the installation process of MongoDB.

\textbf{No Source} \textcolor{purple}{\faTimesCircle} Despite the helpful role of sources in supporting participants to overcome challenges, some participants chose not to use external sources for specific challenges and strategies. This was particularly perceived when dealing with the Lack of Installation Progress Feedback challenge. Regarding strategies, this approach of not consulting sources was observed across six strategies, including Giving Up, Resetting the Environment, Random Actions, Ignoring, and Waiting. These strategies were predominantly employed during the installation processes of Flutter, MongoDB, and Jupyter Notebook.

\section{Discussion}

In this section, we discuss our study's findings by comparing the results from the installation sessions and the validation survey, and by linking them to prior work in the literature. We also discuss the role of sources in the strategies applied when challenges were faced, highlighting both commonalities and discrepancies.

\subsection{Think-aloud Sessions and Validation Survey}

Regarding our validation survey, some of the most common challenges identified in the think-aloud sessions were also widely reported by survey participants. For instance, the challenge Unclear Instructions was encountered by 10 participants during the think-aloud sessions and was also highly reported in the survey, with more than half of participants indicating they have faced this challenge in the past. This aligns with prior studies reporting that faulty and inappropriate installation instructions \citep{aghajani2020software} are among the most common issues faced by developers. Similarly, Dysfunctional Instructions was observed in 10 sessions and was reported by almost half of the survey participants. These results are consistent with findings \citep{mirhosseini2020docable} that show a significant portion of installation tutorials cannot be executed successfully, even with manual intervention.

However, not all challenges and strategies were equally reflected in the survey results. For example, Complicated Installation Process and Lack of Installation Progress Feedback were faced by multiple participants during the think-aloud sessions but were less frequently reported in the survey, with 52 and 29 participants out of 144 selecting them, respectively. This could be due to participants not recalling how they responded to a lack of installation progress feedback events in the past. Furthermore, as this challenge primarily occurred when using the terminal, it is possible that many participants had limited experience with installing tools via the command line. Such installation challenges, particularly those related to setting up the local environment, have also been identified in prior research as significant barriers for newcomers and students working with OSS tools \citep{steinmacher2015systematic, pinto2019training}. Conversely, some challenges were not as frequently encountered during the think-aloud sessions but were commonly reported in the survey, such as Information Overload, Account Required, and Version Incompatibility. Such issues have been previously identified as typical points of friction in installation workflows and configuration processes \citep{aghajani2020software, toth2006experiences}.

In terms of strategies, Searching for Different Documentation and Reading Instructions Carefully were confirmed as common approaches by both the think-aloud participants and the survey respondents. Searching for Different Documentation was applied in 15 think-aloud sessions and was reported as a common strategy by 95 survey participants. This reflects previous findings that developers often consult non-official sources like Stack Overflow and blog posts when official documentation falls short \citep{treude2018does}. Likewise, Reading Instructions Carefully was observed in 14 sessions and was confirmed by more than half of the survey respondents. This is related to work highlighting the importance of clarity and structure in documentation for improving usability \citep{gao2023evaluating}.

In contrast, some strategies were not as widely reported in the survey. For example, Reformulating the Query was applied in 10 installation sessions, but was only reported by 40 survey participants. One possible explanation is that, during the think-aloud sessions, before reformulating the query, participants typically read instructions from one or two sources. Survey participants may have preferred to browse multiple links before considering adding more details to their search query. Another possible explanation is that survey participants may have considered reformulation as an inherent part of the web searching process. This is in line with prior studies noting the difficulty of crafting precise search queries when dealing with fragmented and inconsistent documentation \citep{parnin2011measuring}. Similarly, some strategies that were not as frequently applied in the think-aloud sessions were highly recognised in the survey. For instance, Reading Help or Feedback in Terminal and Reinstalling were not frequently observed during the sessions, but were selected by more than half of the survey participants. The findings of our study indicate that successful installation depends on tool and documentation quality, as well as the effectiveness with which developers employ strategies.

Based on our findings, successful installation depends on tool quality (e.g., Complicated Installation Process, Version Incompatibility) and documentation quality (e.g., Unclear Instructions, Dysfunctional Instructions, Poorly-presented Information) as well as the effectiveness with which developers employ strategies, such as Searching for Different Documentation, Reading Instructions Carefully, or Web Searching. These findings not only confirm but also extend prior research by providing observational evidence of how these challenges and strategies manifest in real-time installation contexts for novice developers.

\subsection{Sources}

In the category of Information Quality/Value challenges, participants using the Searching for Different Documentation strategy consulted different sources depending on the specific sub-challenge. For Unclear Instructions, participants primarily relied on textual documentation. This was because they initially consulted the official documentation, found it unclear, and then turned to alternative sources, still in text format, such as installation guides, how-to tutorials, and similar resources.

In contrast, for the Dysfunctional Instructions challenge, which typically emerged while participants were actively following official instructions and encountered steps that did not work, they tended to seek help from individuals who had experienced the same issues (community knowledge) or turned to video tutorials as a more user-friendly and visually guided format. This contrast highlights that community knowledge and video tutorials were much more frequently consulted in response to Dysfunctional Instructions than Unclear Instructions.

In the category of Information Seeking challenges, installation instructions emerged as a widely consulted source across all strategies. Notably, the Terminal was heavily referenced in the Information Overload challenge, as many of these cases stemmed from the overwhelming volume of information displayed in the terminal. A similar pattern appeared in the Error Message challenge, where participants often encountered issues in the terminal environment. Consequently, most strategies addressing this challenge relied on the terminal itself or involved no source at all. Within the category of Installation Process challenges, the challenges Complicated Installation Process and Lack of Installation Progress Feedback showed clear differences in the sources consulted. Strategies for Complicated Installation Process primarily relied on community knowledge, while Lack of Installation Progress Feedback more frequently involved video tutorials, the terminal, or no source at all.

For the challenges Account Required and Version Incompatibility in Tool Characteristics, different sources were associated with each. In Account Required, the only source consulted was the tool’s homepage. By contrast, strategies for Version Incompatibility involved a wider range of sources, including installation instructions, getting started guides, and community knowledge. Interestingly, within each challenge, the sources consulted were quite consistent across the various strategies applied.

\section{Implications}


Looking at the challenges (see Table \ref{tab:challenges}), it is noticeable that some challenges occurred more frequently than others, hence they had more strategies applied (see Table \ref{tab:sources}). For instance, the challenges \textit{Lack of Installation Progress Feedback}, \textit{Unclear Instructions} and \textit{Dysfunctional Instructions} were the challenges with the highest number of strategies applied. Regarding the strategies, we notice that the strategy \textit{Searching for Different Documentation} stands out, being applied in nine out of the eleven challenges. Implying that, at some point, all participants opted to consult a non-official source in order to find installation instructions, suggesting that official sources are often unreliable. In this section, we outline practical recommendations for tool vendors, tool users, and researchers derived from our findings.


\subsection{Tool Vendors}

\textbf{\textit{Progress Feedback}: }Participants in our study reported difficulties due to insufficient progress feedback, as this challenge occurred in nine sessions. These concerns included not knowing the number of completed and remaining steps, as well as uncertainty about the success of the installation after following all available instructions. The latter issue led to confusion among several participants. Consequently, we advise tool vendors to incorporate progress feedback throughout the installation process, with particular emphasis on confirming the success of the installation at its conclusion. This can be achieved, for example, by providing a test case for users to execute.


\textbf{\textit{Download Link}:} Interestingly, participants experienced difficulties locating download links, a challenge frequently observed in think-aloud sessions, as many expressed frustration over the lack of a straightforward way to install some tools. Although tool vendors may have intentionally designed it this way to encourage potential users to review the documentation before downloading tools, we observed that participants tended to be impatient and anticipated finding download links in more prominent locations.

\textbf{\textit{Dependency Awareness}:} Participants expressed frustration when they perceived that the installation of one tool led to the installation of multiple tools. Unexpected dependencies arose in multiple think-aloud sessions, and 52 survey participants also reported this as a challenge. Although prerequisites and dependencies are often unavoidable, we recommend that tool vendors clearly communicate to potential users from the outset of the installation process that multiple tools will be involved. Additionally, vendors should consider bundling dependencies whenever possible to streamline the installation experience.


\textbf{\textit{Prerequisites}:} To prevent situations where users only discover partway through the installation that their setup does not meet the requirements for a tool, such as due to hardware constraints, prerequisites should be clearly documented. To help users determine whether their setup meets the requirements, tool vendors could provide automated checks as the initial step in the installation process.

\subsection{Tool Users}

\textbf{\textit{Rephrasing Queries}:} In the majority of sessions where participants used rephrasing, this strategy proved successful. Therefore, we recommend that users experiencing challenges during tool installation add more terms related to their specific scenario or context to their initial query. For example, adding keywords such as ``Tool name'' with ``Installation instructions'' or ``Download'' with the operating system name may lead to more successful search results.

\textbf{\textit{Alternative Sources}:} Official documentation frequently lacks guidance on recovering from error scenarios. However, other sources like Stack Overflow are well suited for providing this type of information, considering both participants in the sessions and in the survey perceived this as a useful strategy. Our study revealed that most participants who relied exclusively on official sources were unsuccessful. Therefore, we recommend that potential tool users explore additional sources while remaining aware that they may be less authoritative.

\subsection{Researchers}

\textbf{\textit{Mapping Installation Processes}:} While this study identified challenges and strategies associated with the installation of software development tools, along with the connections between them, future research should focus on obtaining a more canonical understanding of the steps required to install tools, e.g., through additional empirical studies. Our study suggests that some tools had multiple installation paths, and documentation was inconsistent across different sources. A higher-level understanding, such as one codified in an overarching process, could facilitate more systematic research and tool building aimed at addressing the inherent challenges in this field, as well as help organise and consolidate knowledge within the domain.

\textbf{\textit{Automated Techniques for Tool Vendors}:} Our research presents numerous opportunities for future work on the development of automated methods to address the identified challenges. From the tool vendors' perspective, these automated techniques could focus on tasks such as eliminating duplicated instructions, automatically providing progress feedback, automatically testing instructions, augmenting official documentation with relevant information from other sources when necessary, and reducing information overload, among other examples, which were common challenges reported in our study.

\textbf{\textit{Automated Techniques for Tool Users}: }Likewise, there are numerous opportunities to create automated techniques to assist potential tool users, such as query reformulation methods tailored to the context of tool installation, and tools that evaluate a user's environment to determine whether it meets the requirements of a particular tool and its dependencies.

\section{Threats to Validity}

\textit{External Validity. } In our study, we chose to use seven software development tools that cover a wide range of functionalities and purposes, including data management tools (e.g., MongoDB) and machine learning tools (e.g., TensorFlow). Although we believe seven tools to be a sufficient number, we acknowledge that we cannot generalise the results to all software development tools in the market.  We also acknowledge that the identified challenges and strategies might not necessarily be specific to software development tools. Increasing the number of tools could potentially help us identify new challenges and strategies. 


We recognise a potential self-selection bias, as participants voluntarily chose to take part in this study. However, we do not see this as a major issue, given that the study was only advertised to students enrolled in Information Technology subjects. Hence, this study could also be replicated with different populations (e.g., experienced developers) to potentially provide further empirical evidence and compare the challenges of installing software development tools from the perspective of senior developers. Regarding the validation survey, we acknowledge that one challenge and three strategies were not displayed to participants as the survey was conducted before the additional think-aloud sessions, during which these new challenges and strategies were identified. While these were not validated, we believe this caused minimal impact in the results considering they occurred in only a few sessions. However, they may still reflect meaningful experiences for some participants. Future work could explore their prevalence and relevance in broader contexts.

\textit{Internal Validity. } Although the number of tools installed in one session did not exceed three, and each session lasted no more than one hour, we recognise that allowing participants to install multiple tools could have either benefited them due to a training effect or negatively impacted them due to fatigue or frustration if they were unable to successfully install a tool. External factors such as the limited time available for the think-aloud sessions, stress, fatigue, and the stability of the internet connection may also have influenced participants' general experiences during the sessions. Additionally, we acknowledge that the Hawthorne effect \citep{wickstrom2000hawthorne} --- where individuals alter their behaviour due to the awareness of being observed --- may have influenced participants' behaviour, particularly at the beginning of the sessions when they might have been distracted by the observer's presence. However, we do not consider this a significant issue, as participants quickly settled into the task of installing the tools.

We asked participants to use their own machines during the think-aloud sessions. While this decision was made to maintain a realistic setting, we acknowledge that it resulted in a loss of control over pre-installed packages on their devices. Besides, we also recognise that using a clean machine, such as when starting a new job, can also represent a realistic scenario for novice developers. Future work could further investigate the generalisability of our findings to the scenario of using clean machines. In addition, we acknowledge that some participants may have found it easier to install certain tools due to their prior experience with software development tools, while others may have faced more challenges because of their lack of previous experience. However, we believe this was not a significant issue, as participants were novice developers, and challenges were encountered during the installation of most tools, regardless of their prior experience.

Similarly, due to minor adjustments to the study protocol over the course of the study, most but not all participants were asked about their prior experience with the tools they were going to install. Hence, we cannot confirm this information for all participants. In 24 sessions, two participants installing MongoDB and TensorFlow, respectively, were not asked this question. Among those who were asked, one participant had previously installed Flutter, two had installed Jupyter Notebook, two had installed PyTorch, one had installed TensorFlow, and two had installed Node.js. However, none of the tools were already installed on participants' machines, and there was no indication that they had significant experience with these tools. Still, we acknowledge that prior experience or knowledge of the tools may have influenced the results. To explore this, we performed a light-weight comparison between participants with and without prior installation experience. In most cases, participants with experience faced similar challenges and applied comparable strategies to those without experience. In some instances, experienced participants encountered additional or different challenges, but no clear pattern of reduced difficulty or markedly different behaviour emerged. This suggests that prior experience had limited influence on the nature of the challenges encountered or the strategies employed.

While we observed recurring patterns across participants, we acknowledge that including additional participants did lead to the identification of one new challenge and three new strategies. This suggests that although many themes were already emerging, further participation continued to enrich our findings. Furthermore, refining and clarifying the names and descriptions of challenges and strategies improved the clarity of our presentation; however, the coding process itself may be subject to subjective interpretation, which represents a potential threat to validity.

\textit{Construct Validity. }The majority of our study's participants were Master's students who, as mentioned in Subsection 3.1, have been demonstrated to be an adequate proxy for professionals, such as junior developers, in empirical studies. Additionally, our findings are constrained to early-career stage software developers, who may not be familiar with many programming languages, frameworks, and tools.

We acknowledge that participants may have had time to prepare for the installation session, as they pre-selected the tools they would be willing to install before the scheduling of the think-aloud sessions. Also, we recognise that the operating system and the environment used by the participants, and their experience with it could have influenced the challenges they encountered during the installation process. Regarding the validation survey, we acknowledge the possibility of recall bias, as participants may not fully remember the exact challenges they encountered or the strategies they applied when installing software development tools. This is especially relevant since we did not ask them immediately after completing an installation. However, we do not consider this a major issue, given that we conducted think-aloud sessions.

\textit{Conclusion Validity. } Our findings are based on data collected from 18 participants, which we believe to be an adequate sample size. This is because during the think-aloud sessions, we were able to gather more than 500 minutes of video recordings and 187 pages of transcripts. Although we consider this sample size sufficient, involving a larger number of participants could present opportunities to identify new challenges and strategies, as well as validate those we have already discovered. Considering that we distributed the validation survey before conducting the additional think-aloud sessions, new challenges and strategies emerged.

\section{Conclusion}
\label{sec:conclusion}

This research aimed to explore the challenges novice developers face during the software development tool installation process and to discuss the strategies they apply to overcome these obstacles, along with the type of sources they consult when applying these strategies. By analysing data from 18 participants and 24 think-aloud sessions, we identified that novice developers frequently encounter challenges such as unclear or dysfunctional installation instructions, complicated installation steps, version incompatibility, and insufficient installation progress feedback.

Our study also revealed that participants applied strategies such as rephrasing search queries, reading instructions more carefully, and reinstalling tools. One of the most common and effective strategies was searching for alternative sources of information. These sources included non-official documentation such as Stack Overflow posts, blog tutorials, and YouTube videos, in addition to official documentation.

Our findings highlight that successful tool installation depends not only on the technical design of the tool itself but also on the clarity and accessibility of its documentation, and on users’ ability to locate and interpret diverse sources of help. These insights contribute to a better understanding of real-world installation barriers faced by novices and suggest practical improvements for tool vendors, tool users, and researchers.

Based on the findings of our study, several questions remain unanswered, presenting opportunities for future research. For instance, understanding how developers choose between different strategies and sources during the installation process or how their experience level influences installation outcomes could be explored through controlled experiments. Additionally, the long-term effects of positive versus negative installation experiences could be examined through longitudinal studies. Furthermore, with the rise of Large Language Model applications, it would be valuable to investigate how these tools can assist in software development tool installation, potentially addressing common challenges. This could be studied using similar methods to those used in this study, such as conducting observational think-aloud sessions.

Future research can expand upon our findings by identifying additional challenges and understanding the various strategies software developers implement when installing software development tools in specific domains or industries. Additionally, further studies could validate our findings with a larger sample size and investigate tool characteristics, examining how they relate to the challenges and strategies observed.

\section*{Declarations}
\label{sec:declarations}
\textbf{Funding:} This project was not supported by external funding.

\textbf{Competing Interests:} The authors have no conflicts of interest to declare relevant to this paper's content. The authors declare no competing interests. However, Christoph Treude is a member of the EMSE Editorial Board, which is disclosed here for transparency.

\textbf{Ethics Approval:} This study was approved by the Ethics Committee at The University of Melbourne under ID 24329.

\textbf{Consent:} All participants provided informed consent before taking part in this research.

\textbf{Data Availability:} The transcripts from the think-aloud sessions are available at the following repository on GitHub.\\
\url{https://github.com/larisalerno/EMSEChallengesandStrategies}.

\textbf{Authors’ Contribution:} Larissa Salerno was responsible for data collection, data analysis, and manuscript writing. Christoph Treude and Patanamon Thongtatunam contributed to data analysis and manuscript writing.







\bibliography{references}   


\end{document}